\documentclass[acmlarge,nonacm]{acmart}
\AtBeginDocument{}
\setcopyright{none}
\usepackage{tabularx}
\usepackage{threeparttable,enumitem,array}
\renewcommand{\arraystretch}{1.15}

\AtBeginDocument{

}

\usepackage{xltabular}

\begin{document}
\title[Is That Clear?]{Is That Clear? The Impacts of Clarifying Questions on Human-LLM Interaction}
\author{Chengbo Zheng}
\affiliation{%
  \institution{University of Queensland}
  \city{Brisbane}
  \country{Australia}}
\email{chbozheng@gmail.com}

\author{Yujie Zheng}
\affiliation{%
  \institution{Tongji University}
  \city{Shanghai}
  \country{China}}

\author{Qian Zhu}
\affiliation{%
  \institution{Renmin University of China}
  \city{Beijing}
  \country{China}}

\author{Shuai Ma}
\affiliation{%
  \institution{Institute of Software, Chinese Academy of Sciences}
  \city{Beijing}
  \country{China}}

\renewcommand{\shortauthors}{Zheng et al.}
\begin{abstract}
LLM assistants increasingly ask clarifying questions (CQs) to elicit preferences and tailor recommendations. Although prior work emphasizes improvements in answer quality and user experience, how CQs shape users’ subsequent engagement and reliance remains unclear. We conducted a controlled experiment (N=180) across six everyday decision scenarios, comparing human-LLM interaction with and without CQs. We further manipulated whether responses violated explicit constraints and whether violations occurred in the sub-decisions the CQs addressed, while preserving elicited preferences. CQs increased perceived understanding and satisfaction, yet participants spent less time reviewing and evaluating responses and accepted more responses unchanged, including erroneous ones. Within the CQ condition, participants more often accepted responses containing CQ-adjacent violations than nonadjacent violations. These findings demonstrate that LLM's proactive CQs shape subsequent engagement and reliance with preference elicitation. We discuss implications for designing and evaluating proactive assistants that support continued user scrutiny of answers alongside personalization.
\end{abstract}


\maketitle
\hypersetup{pdfauthor={Chengbo Zheng, Yujie Zheng, Qian Zhu, Shuai Ma}}

\section{Introduction}

Large language models (LLM) assistants are increasingly used to support everyday decisions, from choosing products to planning travel~\cite{chatterji2025people}.
A commonly found challenge in human-LLM interaction is users' initial requests are often underspecified~\cite{zamfirescu2023johnny}, leaving LLMs sometimes to postulate the user preference by themselves.
To address this challenge, current LLM assistants are beginning to move beyond responding to the information initially provided and proactively
ask clarifying questions (CQs) to elicit more user contexts before offering a
recommendation.\footnote{See, e.g., product documentation from OpenAI
(\url{https://help.openai.com/en/articles/10500283-deep-research-faq}) and Anthropic
(\url{https://platform.claude.com/docs/en/agent-sdk/user-input}), both accessed September 2026.}
There is growing research examining how such CQs supports preference elicitation and improves users' interaction experiences~\cite{li2025eliciting, sayres2026towards}.

A central motivation for CQs is their informational value: eliciting missing context can help assistants produce answers that better match users' needs, thereby improving task outcomes and satisfaction. 
Research on preference elicitation and conversational clarification has investigated these benefits~\cite{li2025eliciting, sayres2026towards}.
This rationale emphasizes improving final human-LLM interaction outcome by improving what the assistant produces. 
Yet, we postulate a complementary possibility coming out of the user side of CQs-based interaction: \textbf{the clarification exchange itself can shape users' perceptions of the LLM assistant, such as believing LLMs have understood them better, which may, in turn, influence whether and how users engage with and rely on its subsequent answer and impact the human-LLM interaction outcomes.}

Grounding theory offers a basis for considering this possibility. 
In human conversation, people use questions, responses, and acknowledgments to establish that they have understood one another well enough to continue~\cite{clark1991grounding}. 
A completed clarification exchange with an LLM may similarly give users a sense that the assistant understands their needs and can provide a suitable answer. 
This impression may extend beyond the particular preferences discussed, reducing the perceived need to scrutinize the answer or initiate further interaction. 
Conversely, if clarification draws attention to the broader topics discussed, it could encourage closer examination of related content, a possibility based on research on relevance-guided text processing~\cite{mccrudden2007relevance}. 
Whether and how the potential effects might take place is hard to predict and is particularly consequential while LLM responses can contain errors, including recommendations that violate users' stated requirements~\cite{huang2025survey, niu2024ragtruth}.
Yet how CQs shape users' subsequent engagement and reliance of LLMs remains insufficiently understood.

To this end, in this paper we study two questions:

\begin{itemize}
    \item [\textbf{RQ1}] How do CQ affect users' subsequent engagement with, reliance on, and critical evaluation of LLM responses?
    \item [\textbf{RQ2}] When CQs are asked, do reliance and critical evaluation differ between violations in the sub-decisions the CQs addressed (CQ-adjacent) and the one they did not (CQ-nonadjacent)?
\end{itemize}

To address these questions, we conducted a controlled online experiment in everyday decision scenarios (N = 180).
Participants either received the LLM response directly (w/o CQ) or first answered two CQs (w/ CQ). 
Each participant completed three scenarios.
Each scenario further comprised three sub-decisions, each pairing a hard constraint with an unspecified preference. 
One response satisfied all constraints, and two each contained a single planted constraint violation. 
Within the w/ CQ condition, one violation occurred in a sub-decision a CQ had addressed (CQ-adjacent) and the other in the sub-decision no CQ addressed (CQ-nonadjacent). 
In both cases the LLM response preserved the preferences elicited by the CQs, so the violation concerned a hard constraint rather than what the participant had just said, and the violating response remained as personalized as the correct one. 
We therefore study CQs as users encounter them in realistic human-LLM interaction scenarios, as a technique that combines the question interaction with a LLM response shaped by the answers to CQs.
Our study results show that participants who answered CQs spent less time reviewing and evaluating responses, wrote fewer follow-up requests, and more often accepted responses unchanged, including responses containing a violation, while reporting higher perceived alignment and satisfaction and no change in general trust. 
Within the w/ CQ condition, participants more often accepted a CQ-adjacent violating paragraph than a CQ-nonadjacent one, and less often named the violation in their requests. 
CQs thus not only let the assistant give a better answer, but also changed how participants subsequently engaged with it.

In summary, our contributions are threefold:

\begin{itemize}
    \item We found that pre-answer CQs reduce subsequent engagement and increase unchanged acceptance of LLM responses, including erroneous ones, without changing general trust.
    \item We found that when encountering CQs, LLM responses with violations in sub-decisions the CQ asked about are more likely accepted than those with violations elsewhere.
    \item Implications for the design and evaluation of proactive clarification that account for users' subsequent engagement alongside answer quality and user satisfaction.
\end{itemize}

\section{Related Work}

\subsection{Clarifying Questions in Conversational Systems}

Clarification addresses a recurring problem in conversational systems: users' requests may omit information needed to provide useful assistance. 
Research in conversational search shows that well-designed CQs can help users articulate comparison criteria and improve user satisfaction~\cite{bondarenko2022user, zou2023users, zamani2020generating}.
More recently, LLMs are playing more a central role in empowering conversational systems, which bring up challenges for users to clearly articulate their intents and align the LLM answers with user contexts~\cite{subramonyam2024bridging, tankelevitch2024metacognitive}.
Using CQs to elicit user contexts before giving the final answer thus become a promising direction to fill the intent gap~\cite{li2025eliciting}.
Recent HCI research has adopted the CQs as a pre-answer interaction mechanism in tasks such as data wrangling~\cite{chen2025dango}, document generation~\cite{tix2025follow}, literature synthesis~\cite{zheng2024disciplink} and health conversations~\cite{sayres2026towards}.
This line of research often reports improvements to LLM answers thus improving task outcomes~\cite{chen2025dango}, or higher user preferences of the outputs and perceived helpfulness~\cite{tix2025follow, sayres2026towards}.

The benefits of CQ to user experience depend on how the questions are designed. 
In information searching, high-quality CQs encourage in-depth user exploration and improve search performance while low-quality ones hinder the search \cite{zou2023users}.
\citet{hu2024designing} found for follow-up questions aiming for better clarification in conversational agent-based interviews, there are trade-offs among question relevance, the richness of elicited information, and the effort required to respond.
Besides, a recent study found that while CQs can increase user trust and confidence in conversational systems, greater cognitive effort can be imposed~\cite{jung2026ask}.
Recent AI research has devoted to develop the capabilities needed to high quality CQs, including identifying ambiguity and generating relevant questions~\cite{zhang2024clamber}, deciding when clarification warrants additional interaction~\cite{zhang2025clarify}, and eliciting preferences that users may struggle to specify unaided~\cite{li2025eliciting}. 

In summary, existing research has primarily evaluated CQs in terms of elicitation results with CQs, resulting AI output quality and user experience. 
However, how the CQ shapes how users scrutinize subsequent AI answers and how they further engage and rely on them remains insufficiently understood, particularly when those answers contain errors.

\subsection{Conversational Grounding and Perceived Understanding}
\label{sec:grounding}
Conversational grounding describes how people in conversation establish that they have understood one another sufficiently for their current purposes~\cite{clark1991grounding}.
Clarification is one of the grounding acts: asking a question, receiving an answer, and acknowledging it gives both parties evidence that common ground has been reached~\cite{clark1991grounding, benotti2021grounding}.
Previous research has found that it is challenging for LLM assistants to  perform such grounding act well.
\citet{shaikh2024grounding} found that LLMs produce far fewer grounding acts than humans and tend to proceed as if common ground were already established.
CQs, which are system-initiated grounding moves, therefore work as a critical complementary to LLM for producing grounded response.

However, it is critical to note that grounding acts do more than transfer information.
The conversational agents' behavior intended for grounding also shapes how users judge the agents.
Grounding behaviors with high initiative and transparency, such as explaining understanding failure and providing actionable options, reduce user perceived effort and reinforce user trust that the agent is actively processing their input~\cite{ashktorab2019resilient}.
Besides, AI responses can make people feel heard and understood, and this effect can depend on how the interaction is framed rather than only on the content of the answer~\cite{yin2024ai}.
Interactions around CQs may thus raise users' sense that the assistant has understood them, independent of whether the provided information is correct.

This difference between feeling understood and receiving a correct answer matters.
Grounding establishes understanding that is sufficient for the in-situ communication purpose, not that the subsequent parts are right~\cite{clark1991grounding}.
Consistent with this, \citet{sharma2026feedback} identify common-ground failures and difficulties in verifying outputs as separate barriers in conversational agents.
As answering CQs may raise perceived grounding, it may also lower user efforts to verify the subsequent answer, even where errors remain.
Examining CQs beyond their elicitation value is therefore necessary to understand how their impacts to human-LLM interaction.

\subsection{Human–AI Decision Making with Imperfect AI Advice}

Research on human–AI decision making distinguishes encouraging reliance from helping users rely appropriately. 
Explanations can increase acceptance of both correct and incorrect recommendations without improving complementary team performance~\cite{bansal2021does}, while interventions that encourage deliberation can reduce overreliance despite receiving less favorable subjective evaluations~\cite{buccinca2021trust,ma2025towards}. 
These outcomes depend partly on the effort and incentives involved in evaluating advice.
Making verification more worthwhile can reduce overreliance~\cite{vasconcelos2023explanations}, and task characteristics can change reliance without corresponding changes in reported trust~\cite{salimzadeh2024dealing}.
Subjective user experience and appropriate behavioral reliance are therefore related but can diverge.

In conversational LLM systems, the presentation of advice also shapes acceptance and further inquiry. 
For example, first-person uncertainty expression can decrease user reliance on LLM answers~\cite{kim2024m}.
Similar to previous human-AI decision research, \citet{kim2025fostering} found that explanation can increase user reliance, while sources can allow users to reject incorrect LLMs answers more.
They also found that explanations and sources can also reduce users' willingness to ask follow-up questions~\cite{kim2025fostering}.
Moreover, questions explicitly designed to prompt user reasoning with LLM answers can improve users' discernment of flawed claims~\cite{danry2023don}.
Note that such questions are different from CQs studied in this paper, which aims to elicit user preferences before an answer is given.
Building on this work, we examine how CQs preceding an answer affects subsequent reading, reliance, and follow-up engagements.

\section{Study Design and Methodology}

\begin{figure}
    \centering
    \includegraphics[width=1\linewidth]{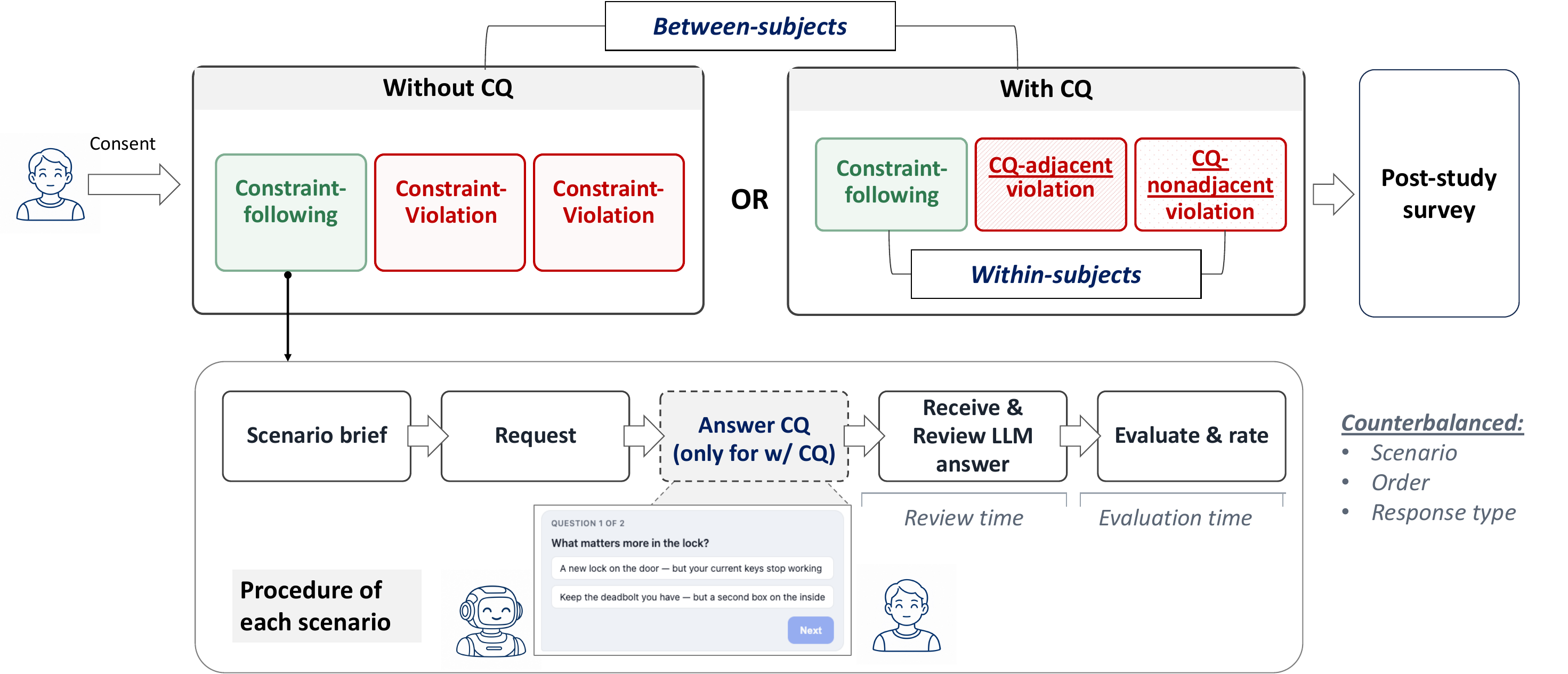}
    \caption{Overview of the study design and procedure. Participants were assigned to either the w/o-CQ or w/-CQ condition and completed three task scenarios. All participants received one constraint-following LLM response and two responses containing a single constraint violation. In the w/-CQ condition, one violation occurred in a sub-decision addressed by a CQ (CQ-adjacent), whereas the other occurred in a sub-decision not addressed by a CQ (CQ-nonadjacent).}
    \label{fig:procedure}
\end{figure}

\begin{figure}
    \centering
    \includegraphics[width=1\linewidth]{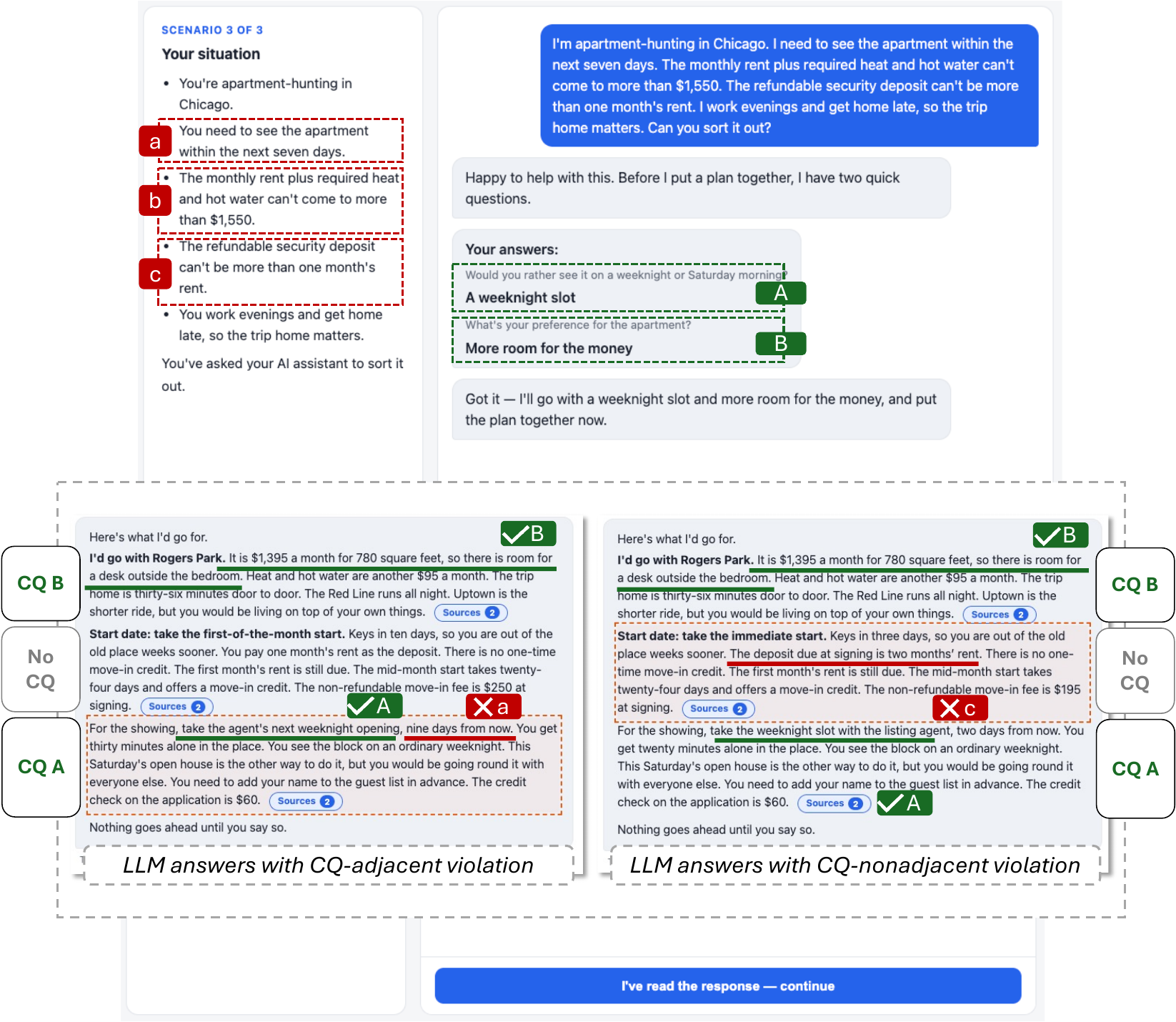}
    \caption{Study interface in the w/ CQ condition, illustrating violation adjacency. The scenario brief states three constraints (a--c) and the assistant asks two CQs (A, B). Each paragraph of the response is tagged with whether it is relevant to CQ (\checkmark{} marks the elicited preference being followed). Left: a CQ-adjacent violation, where the viewing paragraph follows preference A but breaks constraint a. Right: a CQ-nonadjacent violation, where the lease paragraph, which no CQ addressed, breaks constraint c.}
    \label{fig:cq-condition}
\end{figure}

We conducted a controlled online experiment in which participants used a conversational AI assistant for three everyday decision scenarios drawn from a pool of six (\autoref{fig:procedure}).
The assistant either responded without asking clarifying questions (CQs) or first asked CQs.
Each participant received one response that satisfied all scenario constraints and two responses each containing one planted mistake.
To control conditions, the LLM Responses in the study were assembled during the study sessions from a pool of paragraphs generated by LLMs in advance.
The study was approved by the institution's ethics committee.

\subsection{Study Conditions}

Whether the LLM assistant asked CQs was manipulated between participants, with two conditions: \textbf{w/ CQ} and \textbf{w/o CQ}.
In the w/ CQ condition, the assistant asked CQs, waiting for participant answers before presenting its response.
In the w/o CQ condition, the assistant proceeded to the response without asking questions.
Both conditions used the same simulated processing interval before response presentation, as described in the Procedure section.
Both conditions used the same scenario pool. Each scenario involved three sub-decisions, each pairing a hard constraint with an unspecified preference (\autoref{sec:decision-scenarios}).

As shown in \autoref{fig:procedure}, participants in both conditions completed three distinct scenarios, receiving one constraint-following response and two responses containing a constraint violation.
Within the w/ CQ condition, these comprised the following three response types:
\begin{itemize}
    \item \textbf{Constraint-following response.} The response satisfied all stated scenario constraints.
    \item \textbf{Response with a CQ-adjacent violation.} The response violated one constraint in a sub-decision addressed by a CQ, while preserving the preference elicited by that question.
    \item \textbf{Response with a CQ-nonadjacent violation.} The response violated one constraint in the sub-decision not addressed by a CQ.
\end{itemize}

Adjacency, referring to whether a violation concerned a sub-decision addressed by a CQ, was designed for within-subjects comparison to address RQ2 (\autoref{fig:cq-condition}).
Both types of violating response preserved the preferences elicited through CQs. 
The factual information in the violating paragraph agreed with the supporting sources, so participants could identify the violation by comparing the response with the scenario brief alone.
In the w/o CQ condition, the two violating responses were not distinguished by adjacency because no CQs were asked.

\subsection{Participants}
Participants were recruited through Prolific\footnote{\url{https://www.prolific.com/}}.
Eligibility required English as a first language, an approval rate above 99\%, and at least 200 previously completed Prolific studies.
188 participants completed the study while eight were excluded for failing attention checks.
Thus in total, we have 180 participants, including 85 female, 91 male, 1 who preferred not to say, and 3 whose sex was not reported, with a mean age of 34.2 years (\textit{SD} = 6.8, range 19 to 48).
Each condition has 90 participants.
Participants received a base payment of \pounds 3 with the median completion time as 17 minutes.
To encourage participants to treat each scenario as their own decision, we told them before the study that careful answers would earn \pounds 0.20.
The bonus was meant to reward engagement, not error detection, as ``error'' was never mentioned in the instruction.
We paid it to every participant who completed all three scenarios and passed the attention checks. 
Most participants used generative AI tools regularly: 91 reported daily use, 56 weekly use, 26 monthly use, and seven rare use.
In addition, 143 reported that they always or often compared options and checked details themselves when working with AI in daily life.

\subsection{Tasks and Materials}

The study materials included six everyday decision scenarios, CQs about user preferences, and prepared LLM responses. 
Below, we first describe how we selected the type of mistake to include in the responses, developed the scenarios and CQs, and constructed responses for the study conditions.

\subsubsection{Planted mistake type.}
In a pilot study, we compared two types of mistakes that require different forms of verification: constraint violations and factual errors.
A constraint violation recommended an arrangement that broke a stated requirement despite accurately reporting source information; a factual error contradicted a supporting source without violating a stated requirement.
These two types cover the two representative ways a LLM response can diverge from its inputs\cite{huang2025survey,niu2024ragtruth}: from what the user asked for (instruction inconsistency) and from what the supporting materials say (context inconsistency).
Also, in users report, constraint violations are the most frequent complaint~\cite{kim2024understanding}.
We recruited 21 participants through Prolific, each of whom evaluated three scenarios: one with a correct response, one with a factual error, and one with a constraint violation.
Six participants identified the constraint violation, whereas only one identified the factual error.
The low detection of factual errors suggested a potential floor effect that could make differences in critical evaluation difficult to observe.
\textbf{The main study therefore used constraint violations only}, allowing participants to identify mistakes by comparing the response with the scenario brief, which are still challenging based on our pilot results.

\subsubsection{Decision scenarios}
\label{sec:decision-scenarios}
We used six everyday decision scenarios covering shopping, travel, housing, transport, and dining.
Three scenarios were adapted from existing benchmarks for LLM agents: earbuds shopping and smart lock installation from ShoppingCOMP~\cite{tou2025shoppingcomp}, and the New York trip from TravelPlanner~\cite{xie2024travelplanner}.
We developed three additional scenarios to broaden the range of decision contexts.

\begin{itemize}
    \item \textbf{Earbuds shopping.} Choose replacement earbuds for daily cycling, with requirements for ambient sound awareness, returns after opening, and USB-C charging.

    \item \textbf{Smart lock installation.} Choose a smart lock and arrange installation and guest access, ensuring keypad entry, installation within one week, and access throughout a neighbor's cat-sitting visits.

    \item \textbf{New York trip.} Plan an overnight visit, including a hotel within budget, lunch available on Saturday, and afternoon activities requiring no advance booking or payment.

    \item \textbf{Apartment hunting.} Choose an apartment in Chicago, considering viewing availability, rent and mandatory utility costs, the security deposit, and travel home after late shifts.

    \item \textbf{Car purchase.} Choose a car and arrange its purchase and handover, requiring delivery within 30 days, running-cost estimates that do not assume home charging, and seller-managed car registration.

    \item \textbf{Group dinner.} Arrange dinner for friends in Chicago, with a per-person budget covering food and mandatory booking fees, separate payments, and seating together at one table.
\end{itemize}

\paragraph{Sub-decisions, Constraints, and preferences.}
Each scenario comprised three sub-decisions, each pairing a \textit{hard constraint} with \textit{an unspecified user preference} that could be elicited through a CQ.
Hard constraints specified requirements that the response had to satisfy, whereas preferences distinguished LLM answer alternatives to the sub-decisions that could all satisfy those requirements.
For example, the installation sub-decision in the smart lock scenario required installation within one week.
The unspecified preference was whether participants would install the lock themselves or have someone install it.
The supporting sources provided options that accommodated either preference while meeting the installation deadline.

\paragraph{Supporting sources.}
For each scenario, we collected real options from retailers, hotels, restaurants, and listing sites.
We used these sources instead of benchmark-provided sandbox data to ground the LLM recommendations in real offerings and realistic purchasing or booking terms.
We instructed LLMs to examine these sources when generating candidate response content.
Each sub-decision paragraph included a \emph{Sources} badge through which participants could consult its supporting information.
All prepared paragraphs accurately represented the relevant source information, including those that recommended arrangements violating a scenario constraint.

\subsubsection{Clarifying Questions}

\paragraph{CQ construction and selection.}
We developed one CQ per sub-decision, yielding 18 CQs across the six scenarios.
Each question elicited a preference that was unspecified in the scenario and could meaningfully affect the recommendation.
Because prior research links CQ quality to users' engagement and satisfaction~\cite{zou2023users}, we screened candidate questions to reduce the influence of irrelevant or poorly formulated CQs.

For each sub-decision, we generated three candidate CQs using OpenAI GPT-5, yielding 54 candidates.
Two researchers independently rated whether the candidates elicited meaningful preferences that could affect the sub-decision, and we selected the question with the highest mean rating for each sub-decision.
Participants also evaluated the selected CQs during the main study, as described in \autoref{sec:cq-evaluations}.

\paragraph{Preference options of CQs.}
Each CQ offered two options, each describing a benefit and a downside to make the trade-off explicit.
Neither option was inherently correct, and neither altered the hard constraint.
This design allowed participants to express personal priorities among feasible alternatives.
In the w/ CQ condition, the LLM assistant asked about two of the three sub-decisions.
Asking two CQs allowed participants to express preferences across multiple aspects of a scenario, while leaving one sub-decision unasked enabled the comparison between CQ-adjacent and CQ-nonadjacent violations.

\subsubsection{LLM Response Construction}
\label{sec:llm-construction}

\begin{figure}
    \centering
    \includegraphics[width=\linewidth]{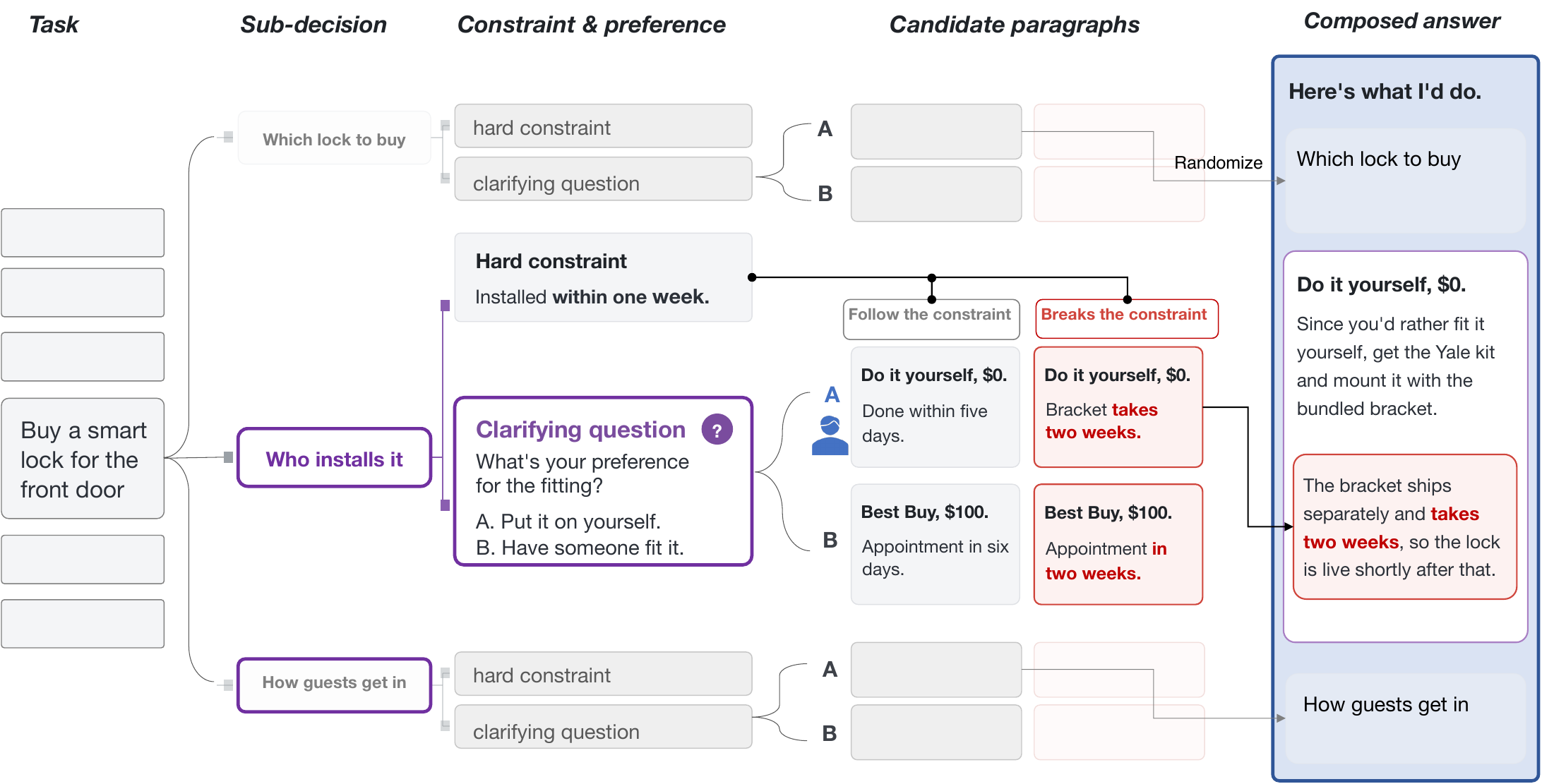}
    \caption{LLM Response construction in the study, illustrated with a CQ-adjacent violation in the smart lock scenario. Four candidate paragraphs were prepared for each sub-decision by crossing two preference options with two constraint statuses. CQ answers determined the preference options for asked sub-decisions; the option for the unasked sub-decision was selected randomly. The condition assignment schedule determined constraint status, and if violated, the violation location. In this illustration, the installation paragraph preserves the preference for self-installation but specifies a two-week delivery time, violating the one-week deadline.}
    \label{fig:task-cq-answer}
\end{figure}

All LLM assistant responses in the study followed the same template: a short opening, one paragraph for each of the three sub-decisions, and a short closing.
Across conditions, the responses are all approximately 200 words.

For each sub-decision of each scenario, we prepared four candidate paragraphs by crossing two dimensions: \underline{preference option} (the two CQ options) and \underline{constraint status} (satisfying or violating the hard constraint; \autoref{fig:task-cq-answer}).
For each preference option, the violating candidate preserved that preference and matched the non-violating candidate in structure and approximate length, but recommended an arrangement that broke the hard constraint.
The two violating candidates for a sub-decision, one per preference option, violated the same constraint in the same way (e.g., a two-week installation timeline against a one-week deadline in \autoref{fig:task-cq-answer}), so each sub-decision carried one designated violation regardless of the elicited preference.

Two researchers independently examined all 72 candidate paragraphs to verify that non-violating candidates satisfied the relevant constraints and that each violating candidate contained exactly one designated constraint violation.

During the study, responses were assembled by selecting one candidate paragraph for each sub-decision (\autoref{fig:task-cq-answer}).
First, the \underline{preference option} was determined by the participant's answer when a CQ was asked, or selected randomly from the two options when no CQ was asked.
This random selection applied to all three sub-decisions in the w/o CQ condition and the one unasked sub-decision in the w/ CQ condition.
Second, the \underline{constraint status} was determined by the assigned response type and violation location.
Constraint-following responses used three non-violating paragraphs; responses containing a violation used one violating paragraph and two non-violating paragraphs.
In the w/ CQ condition, the violating paragraph concerned an asked sub-decision for CQ-adjacent violations and the unasked sub-decision for CQ-nonadjacent violations.
Violation locations followed the counterbalanced schedules described in \autoref{sec:counterbalancing}.

\subsection{Trial Assignment and Counterbalancing}
\label{sec:counterbalancing}

We created 18 schedules to counterbalance scenario selection, presentation order, and response type.
Each schedule specified a participant's three scenarios and their order, which response would satisfy all constraints, the sub-decision containing each of the two violations, and the two sub-decisions to be addressed by CQs in each scenario.
Our study material in total has 18 different constraint violations (six scenarios $\times$ three sub-decisions; see \autoref{sec:llm-construction}).
The schedules were designed to cover all these violations, with each  appeared once as a CQ-adjacent violation and once as a CQ-nonadjacent violation.
Besides, each scenario received a constraint-following response in three schedules and appeared three times in each presentation position  across all response types.
The w/o CQ condition used the same schedules.
The order of the two CQs and the preference options for unasked sub-decisions were randomized within each session.
We used Prolific's TaskFlow to distribute participants evenly across the 18 schedules, with ten participants per schedule (five per condition), yielding a fully counterbalanced sample.
Consequently, each designated violation was presented to 20 participants across the two conditions.
The complete schedules are provided in the Appendix.

\subsection{Procedure}

Participants provided consent and read instructions that included a walkthrough of the interface using a practice example in their assigned condition.
They were instructed to evaluate each response against the scenario's stated requirements while also considering their own preferences.
A single-choice comprehension question checked their understanding of the instruction to use their personal preferences.

\begin{figure}
    \centering
    \includegraphics[width=1\linewidth]{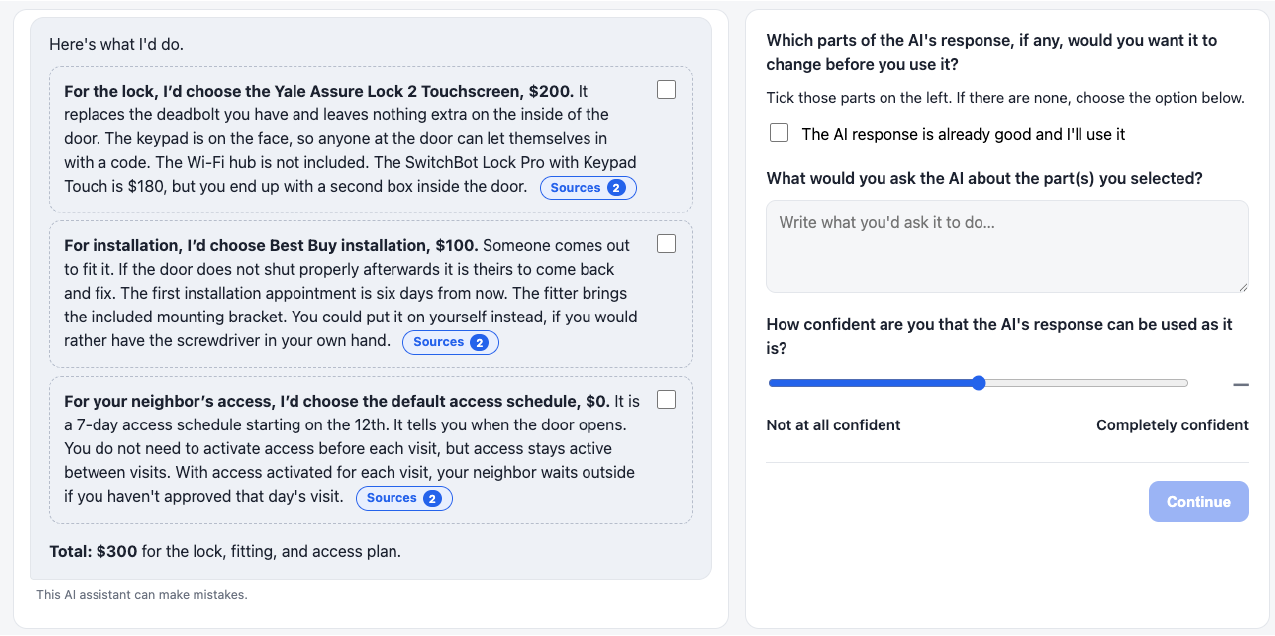}
    \caption{LLM Response evaluation study interface. Participants selected sub-decision paragraphs they wanted changed or indicated that they would use the response unchanged. Those selecting paragraphs entered a written request. Participants then rated their confidence in using the response as presented. The response and its supporting sources remained accessible during evaluation.}
    \label{fig:case-question}
\end{figure}

For each of the three scenarios, participants completed the following steps (\autoref{fig:procedure}):
\begin{itemize}
    \item \textbf{Reading the scenario.} A brief presented the situation and its requirements as bullet points. Participants could proceed after a minimum of 20 seconds reading.
    \item \textbf{Interacting with the assistant.} Participants entered a chat interface with the scenario brief displayed on the left (\autoref{fig:cq-condition}). They pressed \emph{Send} to submit a fixed, prewritten request containing the scenario information. In the w/ CQ condition, they then answered two CQs. After these answers, or immediately after the request in the w/o CQ condition, the interface displayed simulated search and planning traces for approximately 10 seconds before streaming the prepared response. A button for proceeding to the evaluation became available after a minimum review period of 20 seconds~\footnote{Mean review time exceeded 50s in both conditions (see the Results section)}.
    \item \textbf{Completing the post-interaction questionnaire.} A questionnaire panel opened on the right while the response remained visible. On the first page, participants selected any of the three sub-decision paragraphs they wanted changed before using the response, or selected ``\textit{The AI response is already good and I'll use it.}'' (\autoref{fig:case-question}). Participants who selected paragraphs wrote what they would ask the assistant about those parts. They then rated their confidence in using the response unchanged. The second page collected ratings of mental demand and response usefulness. On the third page, participants in the w/ CQ condition rated the usefulness of each CQ. The scales are specified in \autoref{sec:measures}.
\end{itemize}

After all three scenarios, participants completed a post-study questionnaire covering trust, their experience with the assistant, and how they had evaluated its responses.
Next, each of the two planted violations they had encountered was explained.
For each violation, participants reported whether they had noticed that specific issue, suspected a problem, or noticed no problem.
They were informed that these answers would not affect payment.
The study ended with background questions about how often participants checked details themselves and used generative AI tools.

\subsection{Measures}
\label{sec:measures}

We collected behavioral logs, response acceptance and paragraph selections, written requests, and self-reported ratings to examine participants' engagement, reliance, and critical evaluation of LLM responses.
Measures were recorded per scenario, yielding 540 LLM response-level observations, including 360 with a constraint violation in the response.
Here by engagement, we measure the engagement concerning user interaction after the response arrived.
User interaction with CQs was not included as the analysis of the engagement.

\subsubsection{Behavioral Engagement}
We measured time spent reviewing and evaluating responses and consulting sources~\cite{jansen2025user}, analyzing these indicators separately.

\begin{itemize}
\item \textit{Review time}: time from the moment where the LLM assistant had delivered the response to participants confirmed they had read the response and opened the questions to answer. The confirmation button unlocked after 20 seconds, setting a minimum review time of 20 seconds.
\item \textit{Evaluation time}: time from opening the evaluation page to submitting paragraph selections, any written request, and confidence ratings. This page had no minimum time requirement.
\item \textit{Source use}: The number of source openings. Each paragraph had one source tag.
\end{itemize}

\subsubsection{Reliance}
We operationalized reliance as willingness to use an LLM response unchanged, drawing on prior work measuring agreement with AI recommendations~\cite{ma2025towards, ma2023should, kim2025fostering}.
We also assessed confidence in using the response.

\begin{itemize}
\item \textit{Case-level agreement}: whether participants selected ``\textit{The AI response is already good and I'll use it.}'' for the response, coded as 1 for agreement and 0 otherwise.
\item \textit{Paragraph-level agreement}: the proportion of the three paragraphs within each response that participants did not select for changes.
\item \textit{Confidence}: confidence in using the LLM response unchanged, rated from 0 to 100.
\end{itemize}

Participants requests for changes could reflect either a constraint violation or a personal preference, so we did not classify acceptance and revision requests as overreliance or underreliance~\cite{ma2025towards,schemmer2023appropriate}.
We used coding of participants' request in changing the answers to distinguish different reasons.

\subsubsection{Critical Evaluation}
We examined whether participants explicitly described the planted violation in their requests and retrospectively reported noticing it.

\begin{itemize}
\item \textit{Error reporting}: whether the written request correctly described the planted violation (i.e., with the FoundError code, detailed in \autoref{sec:analysis}), regardless of which paragraph was selected. Responses without a written request were coded as no error reporting.
\item \textit{Error awareness}: whether participants retrospectively reported noticing the specific issue, suspecting a problem without identifying it, or noticing no problem.
\end{itemize}

\subsubsection{User Perceptions}

\paragraph{Response evaluations.}
After each scenario, participants rated \textit{usefulness} (``This response gave me useful information for deciding on a final plan'') and \textit{mental demand} (``How mentally demanding was it to evaluate this response?'')~\cite{buccinca2021trust,hart1988development}.
Both used seven-point scales.
Participants completed these ratings only after recording their response decisions.

\paragraph{Assistant perceptions.}
The post-study questionnaire assessed the following aspects of participants' experience, all with seven-point scales.

\begin{itemize}
\item \textit{Trust}: three items from the Short Trust in Automation Scale~\cite{jian2000foundations,mcgrath2025measuring}.
\item \textit{Alignment}: three items assessing how well participants felt that the assistant aligned with their intents. To our knowledge, no established scale directly measures this perception. Based on prior HCI research of the three alignment gulfs identified by \citet{subramonyam2024bridging} and research pursuing a similar measurement goal~\cite{wen2026adaptive}, we therefore included:
``The assistant understood which requirements and preferences mattered to me'' ( \textit{intention gap}); ``I was able to communicate what mattered to me clearly''  (\textit{instruction gap}); and ``The assistant can generate plans reflected what I was looking for'' (\textit{capability gap}).
\item \textit{Satisfaction}: a single item, ``\textit{I was satisfied with this assistant.}''
\item \textit{Evaluation experience}: three item of the perceived breadth of considerations (``\textit{I had many different requirements and preferences to keep track of while evaluating the plans}''), distribution of the attention (``\textit{My attention was spread across several aspects of the plans}''), and whether participants felt they needed to double-check the plans.
\end{itemize}

\paragraph{CQ quality evaluations.}
We assessed how participants experienced the CQs and their general attitudes toward assistants that ask questions before answering.
After each scenario, participants in the w/ CQ condition rated each question's \textit{usefulness} on a 0--10 scale
In the post-study questionnaire, they rated four additional items on seven-point scales of their perceived \textit{relevance}, \textit{preference reflection}, \textit{effort}, and \textit{trust impacts} of the encountered CQs.

\subsection{Analysis}
\label{sec:analysis}

\subsubsection{Quantitative Analysis}

We analyzed the data with mixed-effects models with random intercepts for participants and scenarios:
\begin{equation*}
\texttt{DV \textasciitilde{} CQ + Adjacency + (1|participant) + (1|scenarios)}.
\end{equation*}
\texttt{CQ} was coded TRUE for the w/ CQ condition and FALSE for the w/o CQ condition.
\texttt{Adjacency} was coded $+0.5$ for adjacent errors, $-0.5$ for nonadjacent errors, and 0 for the w/o CQ condition and correct responses.
The CQ coefficient estimated the between-condition contrast, while the adjacency coefficient estimated the contrast between adjacent and nonadjacent errors within the w/ CQ condition.
This single model thus uses all available observations, accounts for participant and case variance, and yields both research focuses directly (i.e., CQ for the between-condition effect and Adjacency for the within-condition contrast).
We applied Holm corrections separately within the measure groups.

We fitted logistic mixed-effects models for binary outcomes, such as case-level agreement.
For continuous outcomes, such as confidence, we fitted linear mixed-effects models.
Scenario order was counterbalanced across participants, so we did not include it in the models reported here. 
Nevertheless, we tested adding order as a covariate and found it does not affect the statistic testing results.
Time measures were log-transformed, with trial position included as an additional covariate.
For the post-study surveys, we compared the two conditions using Welch's $t$ tests as the seven point likert scales were used~\cite{norman2010likert}.

\subsubsection{Request Coding}

In total, we collected 254 written requests from 131 participants.
We analyzed these texts to examine what they sought to follow up with the LLM assistant.
First, two researchers went through the data together and developed a codebook:

\begin{itemize}
\item \textit{FoundError}: correctly identifies the planted violation.
\item \textit{AskVerify}: asks the assistant to check a stated constraint without explicitly asserting a violation.
\item \textit{FalseAlarm}: claims a violation that the response did not contain.
\item \textit{Preference}: requests a different option based on personal preference, not relevant to the scenario constraints.
\item \textit{Understand}: asks for information, reasoning, clarification, or additional options.
\item \textit{NoContent}: contains no substantive request.
\end{itemize}

Two authors coded all 254 texts independently.
Because a text can carry several codes, we computed Cohen's $\kappa$ for each code separately, which was .98 for FoundError, 1.00 for NoContent, .96 for Preference, .94 for Understand, .79 for AskVerify, and .76 for FalseAlarm.
The two coders assigned exactly the same six codes on 94\% of the texts.
They then discussed every text they disagreed on and settled on one set of codes for each, and the analysis uses those agreed codes.

\section{Study Results}

We report the results in the order of the measures.
For each measure, we report the between-condition comparison (RQ1) and then the within-condition adjacency contrast (RQ2).
We then examine the content of follow-up requests and participants' perceived experience.
Unless explicitly described as participant-level summaries or post-study ratings, the results concern individual participants' evaluations of individual scenario responses.

\subsection{Participants Engaged Less with Responses after CQs}
\label{sec:results-engagement}

Participants who answered CQs spent less time reviewing and evaluating responses, both for correct and erroneous responses, while source use did not differ.

\begin{figure}
    \centering
    \includegraphics[width=1\linewidth]{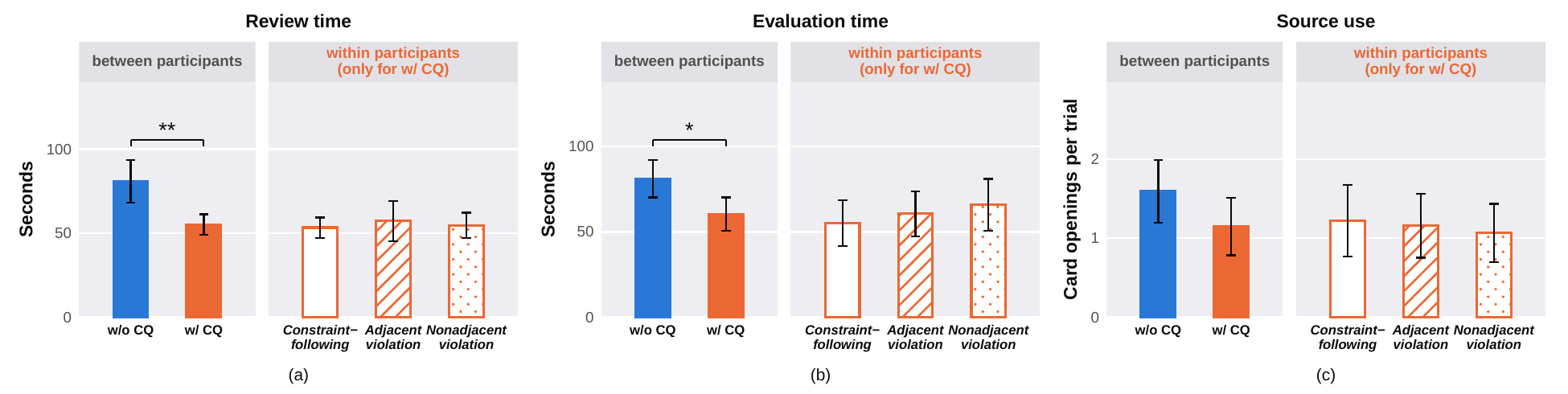}
    \caption{Behavioral engagement by condition and response type: (a) review time, (b) evaluation time, and (c) source use. Within the w/ CQ condition, responses are separated into those that followed the constraints and those containing adjacent or nonadjacent violations.}
    \label{fig:engagement}
\end{figure}

\paragraph{Review Time}
Participants who did not encounter CQs spent a mean of 80.7 seconds ($SD = 60.4$) reviewing the LLM response before opening the evaluation questions.
Participants who encountered CQs spent less time reviewing the response, with a mean of 55.0 seconds ($SD = 29.3$).
In the mixed-effects model, review time in the w/ CQ condition was 0.77 times that in the w/o CQ condition ($p = .003$, Figure~\ref{fig:engagement}a).
Participant-level means showed the same significant difference ($p < .001$, $d = -0.54$).

Because identifying a violation could require additional review, we also examined trials under both conditions with LLM responses without any constraint violations.
Review time remained shorter in the w/ CQ condition for this subset, with a mean of 53.3 seconds ($SD = 28.8$) compared with 75.5 seconds ($SD = 72.0$) in the w/o CQ condition ($p = .007$).
Also, as shown in \autoref{fig:engagement}a, no significant difference is found within the w/ CQ group regarding different types of LLM responses.
Review time did not differ between responses that followed the constraints and responses containing a violation ($p = .94$), or between adjacent and nonadjacent violations ($p = .82$).

\paragraph{Evaluation Time}
Participants in the w/ CQ condition also spent less time completing the evaluation page.
The time from opening the questions to submitting the decision was 0.73 times that in the w/o CQ condition ($p = .024$, Figure~\ref{fig:engagement}b).
In a comparison of participant-level means, participants in the w/o CQ condition took 81.0 seconds on average ($SD = 52.1$), compared with 60.4 seconds ($SD = 46.4$) in the w/ CQ condition ($p = .006$, $d = -0.42$).

\paragraph{Source Use}
Figure~\ref{fig:engagement}c presents the number of source openings per scenario.
Participants in the w/ CQ condition opened sources on 37\% of responses, compared with 46\% in the w/o CQ condition.
They opened a mean of 1.1 source cards per response ($SD = 1.7$), compared with 1.6 ($SD = 1.9$) in the w/o CQ condition.
Neither difference was significant (share of responses, $p = .22$; number of openings, $p = .27$).
\newline

Within the w/ CQ condition, review time, evaluation time, and source use did not differ significantly between LLM responses containing adjacent and nonadjacent violations (all $p \geq .57$).
None of the three measures differed significantly between responses that followed the constraints and responses containing a violation either.

\subsection{Participants Accepted More Responses after CQs}
\label{sec:results-reliance}

We examined case-level agreement (i.e., per scenario participants encountered), paragraph-level agreement, and confidence in using the response unchanged.
For each measure, we first report the between-condition comparison, followed by the available comparisons by response correctness and violation adjacency.

In a nutshell, case-level agreement, paragraph-level agreement, and confidence were all higher after CQs.
Within the CQ condition, all three were higher for responses with CQ-adjacent violations than for those with CQ-nonadjacent violations.

\begin{figure}
    \centering
    \includegraphics[width=1\linewidth]{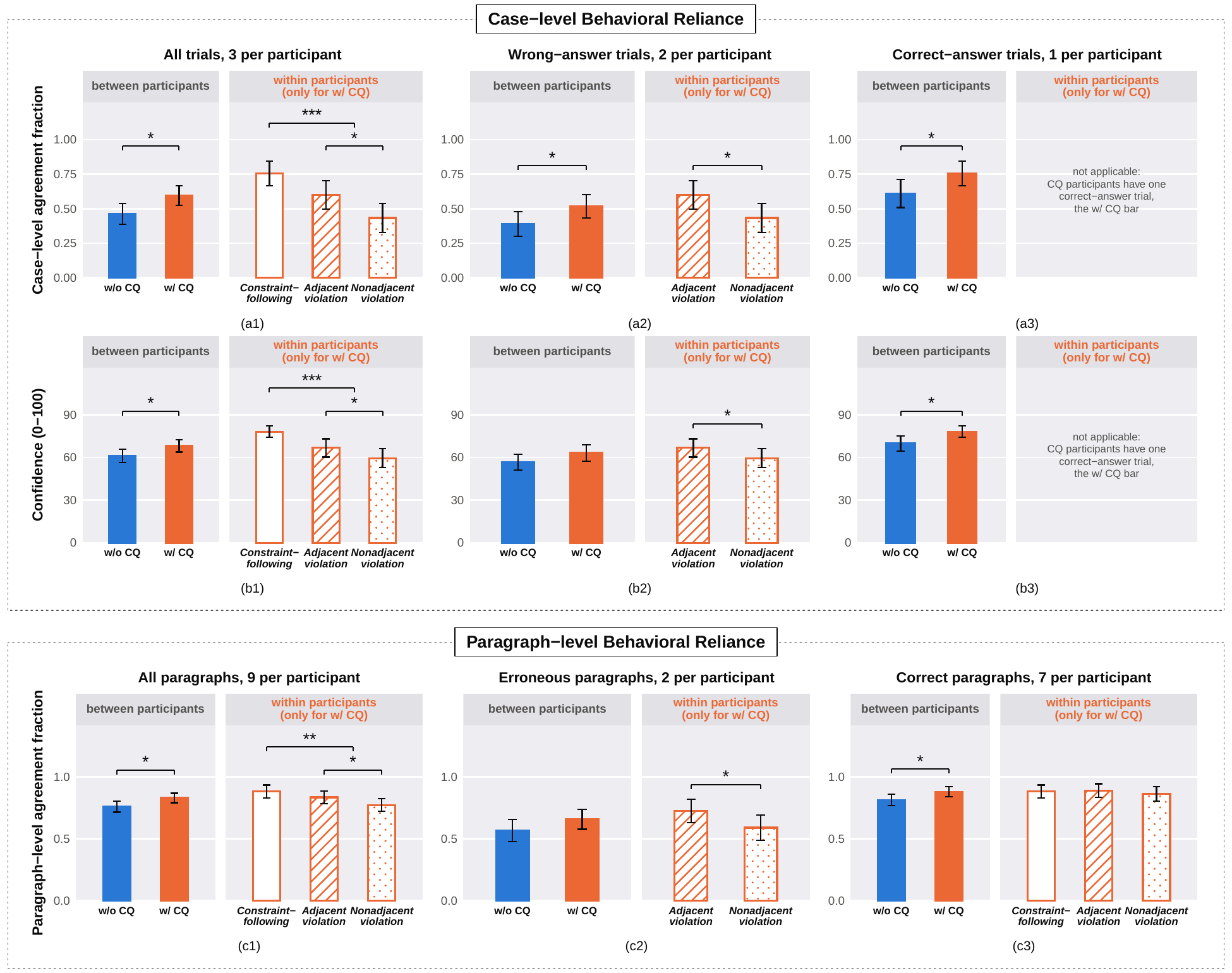}
    \caption{Reliance by the between-subject conditions and the within-subject conditions in the w/ CQ group. Panels (a1--a3) show case-level agreement, (b1--b3) show confidence, and (c1--c3) show paragraph-level agreement. The panels distinguish all responses or paragraphs, those containing violations, and those without violations.}
    \label{fig:behavior}
\end{figure}

\subsubsection{Case-Level Agreement}
\label{sec:results-acceptance}

Participants accepted 46\% of responses unchanged in the w/o CQ condition ($SD = 36$ percentage points across participants), less than the w/ CQ condition with 60\% unchanged ($SD = 34$; Figure~\ref{fig:behavior}a1; $p = .024$, $d = 0.38$).
For trials with \textit{correct} responses, the agreement was 61\% in the w/o CQ condition, lower than 76\% in the w/ CQ condition ($p = .039$, Figure~\ref{fig:behavior}a3).
For trials with \textit{incorrect} responses (combining CQ-adjacent and CQ-nonadjacent violations), agreement was 39\% in the w/o CQ condition and 52\% in the w/ CQ condition ($p = .034$, Figure~\ref{fig:behavior}a2).
In both conditions, participants agreed with correct responses more often than with responses containing a violation ($p < .001$).

Within the w/ CQ condition, agreement was 60\% for responses containing adjacent violations, significantly higher than responses containing nonadjacent violations with 43\% agreement (Figure~\ref{fig:behavior}a2, $p = .019$).

\subsubsection{Paragraph-Level Agreement}
\label{sec:results-paragraph-agreement}

Mean paragraph-level agreement was 0.76 in the w/o CQ condition ($SD = 0.21$ across participants), lower than 0.83 in the w/ CQ condition ($SD = 0.18$; Figure~\ref{fig:behavior}c1, $p = .024$).

For paragraphs containing violations, agreement was 57\% in the w/o CQ condition and 66\% in the w/ CQ condition, but this difference was not significant ($p = .12$, Figure~\ref{fig:behavior}c2).
For paragraphs without violations, agreement was higher in the w/ CQ condition, at 88\% compared with 81\% ($p = .017$, Figure~\ref{fig:behavior}c3).

\label{sec:results-relatedness}
Within the w/ CQ condition, agreement was 72\% for paragraphs containing adjacent violations, higher than 59\% for paragraphs containing nonadjacent violations (Figure~\ref{fig:behavior}c2, $p = .037$). 
A within-subjects paired comparison also showed higher agreement for adjacent violations ($p = .028$, $d_z = 0.24$). 
For paragraphs without violations, agreement did not differ significantly by violation adjacency (89\% vs.\ 86\%, $p = .37$).

\subsubsection{Confidence}

Mean confidence was 61.1 in the w/o CQ condition ($SD = 22.8$ across participants) and 68.1 in the w/ CQ condition ($SD = 20.9$).
Participants in the w/ CQ condition were significantly more confident in using the LLM response unchanged ($p = .032$, Figure~\ref{fig:behavior}b1).
For correct responses, mean confidence was 69.8 ($SD = 25.7$) in the w/o CQ condition, significantly lower than 78.1 ($SD = 19.0$) in the w/ CQ condition ($p = .014$, Figure~\ref{fig:behavior}b3).
For responses containing violations, mean confidence was 56.8 ($SD = 26.3$) in the w/o CQ condition and 63.1 ($SD = 28.1$) in the w/ CQ condition.
The estimated difference was smaller and not significant ($p = .12$, Figure~\ref{fig:behavior}b2).

Within the w/ CQ condition, confidence was 7.2 points higher for responses containing adjacent violations than for responses containing nonadjacent violations ($M = 66.7$, $SD = 30.7$ vs.\ $M = 59.5$, $SD = 32.1$; $p = .037$, Figure~\ref{fig:behavior}b2).

\subsection{CQ-Adjacent Violations Were Less Often Found}
\label{sec:results-critical-evaluation}

Regarding critical evaluation, for between-subject conditions, error reporting and error awareness did not differ significantly. 
Within the CQ condition, participants named CQ-adjacent violations less often than CQ-nonadjacent ones.

\subsubsection{Error Reporting}
Participants named the planted violation in their request on 38\% of erroneous responses in the w/o CQ condition, compared with 30\% in the w/ CQ condition.
This difference was not significant ($OR = 0.54$, $p = .14$).

Within the w/ CQ condition, participants reported the violation on 23\% of responses containing adjacent violations, lower than 37\% for responses containing nonadjacent violations ($p = .019$).
The lower reporting rate for adjacent violations accompanied higher paragraph-level agreement (\autoref{sec:results-relatedness}).
This suggests participants not only accepted CQ-adjacent violating paragraphs more often but also less often identified the violation when they did request changes.

\subsubsection{Error Awareness}

Participants reported noticing the specific issue on 50\% of erroneous responses in the w/o CQ condition and 40\% in the w/ CQ condition, without significant difference ($p = .075$).
Including responses on which participants suspected something was wrong increased these rates to 73\% and 63\%, respectively ($p = .056$).
Within the w/ CQ condition, reported awareness was 37\% for adjacent violations and 43\% for nonadjacent violations, with no significant difference ($p = .31$).
Including suspected responses, the rates were 59\% and 68\% ($p = .19$).

Participants agreed with the violating paragraph on 24\% of responses for which they reported noticing the issue, compared with 92\% of responses for which they reported not noticing it.
Responses on which participants only suspected something was wrong resembled the latter, with 90\% agreement.
Participants named the violation in their request on 72\% of the responses they reported noticing, and on 4\% or fewer of the others.
Table~\ref{table:detect} presents paragraph-level agreement, error reporting, and error awareness together.

\begin{table}
    \centering
    \caption{Agreement with the violating paragraph, error reporting, and retrospective error awareness on the erroneous responses. The upper rows split the responses by conditions. The lower rows split the same responses by what participants reported at the end of the study. Reporting is the portion of responses whose written request named the violation.}
    \label{table:detect}
    \begin{tabular}{lrrrrr}
    \toprule
    Responses & $n$ & Agreement & Reported & Noticed & Noticed or suspected \\
    \midrule
    w/o CQ & 180 & 57\% & 38\% & 50\% & 73\% \\
    w/ CQ & 180 & 66\% & 30\% & 40\% & 63\% \\
    \quad adjacent violation & 90 & 72\% & 23\% & 37\% & 59\% \\
    \quad nonadjacent violation & 90 & 59\% & 37\% & 43\% & 68\% \\
    \midrule
    Reported noticing & 162 & 24\% & 72\% & & \\
    Reported suspecting & 84 & 90\% & 1\% & & \\
    Reported not noticing & 114 & 92\% & 4\% & & \\
    \bottomrule
    \end{tabular}
\end{table}

\subsection{Fewer Follow-up Requests after CQs}
\label{sec:results-requests}

Participants wrote in total 254 requests, with a median length of 18 words.
131 out of 180 participants wrote at least one request, 70 in the w/o CQ condition and 61 in the w/ CQ condition.
With each condition had 270 trials that participants could follow up with requests, participants in the w/o CQ condition wrote a request on 54\% of them, compared with 40\% in the w/ CQ condition, which mirrors the difference in case-level agreement.
Table~\ref{table:requests} shows how often each code appeared, first across all trials and then across the trials that contained a violation.

\textit{FoundError} was the most common code in both conditions, and across all trials its share did not differ significantly (26\% vs.\ 20\%).
The difference lies inside the w/ CQ condition instead.
\textbf{Participants named the violation on 23\% of the trials with a CQ-adjacent violation, against 37\% of the trials with a CQ-nonadjacent violation ($p = .019$)}.
This complements our findings on agreement fractions, on a stricter criterion, as the text has to name the violation and not only mark the paragraph.
\textit{Preference} requests were the clearest between-subject condition difference, falling from w/o CQ's 19\% to w/ CQ's 11\% ($p = .023$).
\textit{Understand} requests were also less common in the w/ CQ condition (16\% vs.\ 11\%), but this difference was not significant.

\begin{table*}
    \centering
    \small
    \caption{Percentage of trials whose participant follow-up request carried each code.
    Tests compare w/ CQ against w/o CQ on the left and CQ-adjacent violation against nonadjacent on the right.}
    \label{table:requests}
    \begin{tabular}{lrr@{\hskip 1.5em}rrr}
    \toprule
     & \multicolumn{2}{c}{All scenario trials} & \multicolumn{3}{c}{Trials containing a violation} \\
    \cmidrule(lr){2-3} \cmidrule(lr){4-6}
    Code & w/o CQ & w/ CQ & w/o CQ & w/ CQ adj.\ viol. & w/ CQ nonadj.\ viol. \\
     & (270) & (270) & (180) & (90) & (90) \\
    \midrule
    Any request & 54\% & 40\%* & 61\% & 40\%* & 57\% \\
    \midrule
    FoundError  & 26\% & 20\%  & 38\% & 23\%* & 37\% \\
    AskVerify   &  3\% &  3\%  &  4\% &  4\%  &  4\% \\
    FalseAlarm  &  1\% &  1\%  &  1\% &  0\%  &  1\% \\
    Preference  & 19\% & 11\%* & 18\% &  9\%  & 13\% \\
    Understand  & 16\% & 11\%  & 14\% & 12\%  & 11\% \\
    \bottomrule
    \end{tabular}
\end{table*}

\subsection{User Perceptions}
\label{sec:results-perceptions}

Participants who answered CQs felt better understood, were more satisfied, and reported less need to double-check, while trust in the assistant did not differ significantly.

\subsubsection{Response Evaluations}
Participants rated the usefulness of the response at a mean of 5.4 in the w/o CQ condition ($SD = 1.0$ across participants) and 5.6 in the w/ CQ condition ($SD = 0.9$), without significant difference ($p = .15$).
Responses containing a violation were rated 0.35 points less useful than correct responses ($p = .004$).
Within the w/ CQ condition, responses containing adjacent violations were rated 0.28 points more useful than responses containing nonadjacent violations ($p = .045$).

Mental demand of evaluating the response was rated at a mean of 3.8 in the w/o CQ condition ($SD = 1.3$) and 3.6 in the w/ CQ condition ($SD = 1.6$), without significant difference ($p = .53$).
Mental demand did not differ by response correctness ($p = .11$) or by violation adjacency ($p = .69$).

\subsubsection{Assistant Perceptions}

\begin{figure}
    \centering
    \includegraphics[width=0.85\linewidth]{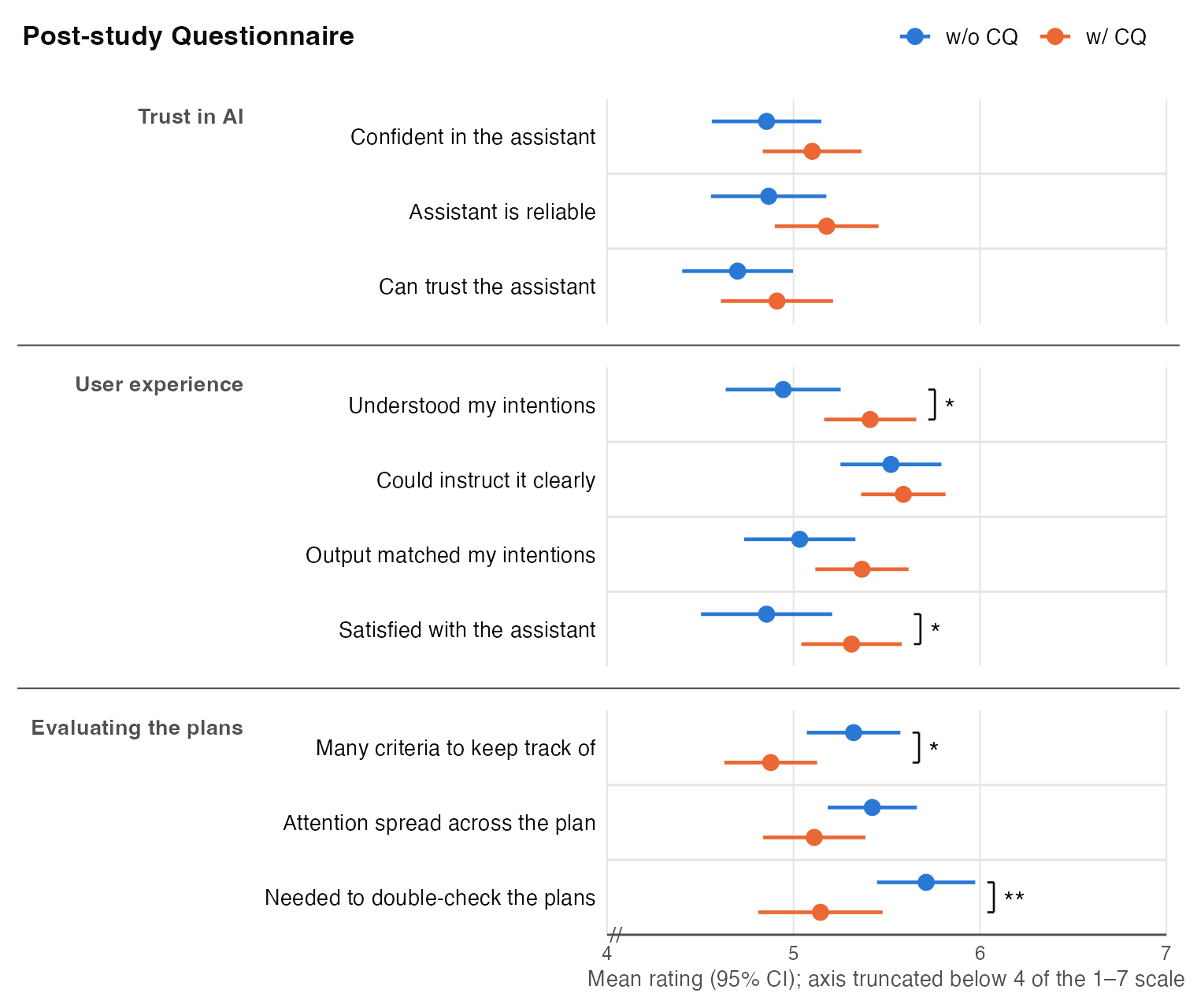}
    \caption{Post-study ratings of trust, alignment, satisfaction, and evaluation experience by condition. Points show mean ratings and intervals show 95\% confidence intervals on seven-point scales.}
    \label{fig:survey}
\end{figure}

\begin{itemize}
    \item \textbf{Trust.} The three trust items did not differ significantly between the w/o and w/ CQ conditions (all $p \geq .14$, Figure~\ref{fig:survey}). 
    \item \textbf{Alignment.} Participants in the w/ CQ condition rated the assistant's understanding of their requirements and preferences higher than those in the w/o CQ condition ($M = 5.4$, $SD = 1.2$ vs.\ $M = 4.9$, $SD = 1.5$; $p = .020$), corresponding to the intention-gap item~\cite{wen2026adaptive, subramonyam2024bridging}.
    The instruction-gap item ($M = 5.6$, $SD = 1.1$ vs.\ $M = 5.5$, $SD = 1.3$; $p = .71$) and the capability-gap item ($M = 5.4$, $SD = 1.2$ vs.\ $M = 5.0$, $SD = 1.4$; $p = .091$), did not differ significantly.
    \item \textbf{Satisfaction.} Participants in the w/ CQ condition were more satisfied with the assistant than those in the w/o CQ condition ($M = 5.3$, $SD = 1.3$ vs.\ $M = 4.9$, $SD = 1.7$; $p = .043$).
    \item \textbf{Evaluation Experience.} Participants in the w/ CQ condition reported fewer criteria to keep track of ($M = 4.9$, $SD = 1.2$ vs.\ $M = 5.3$, $SD = 1.2$; $p = .013$) and less need to double-check the responses ($M = 5.1$, $SD = 1.6$ vs.\ $M = 5.7$, $SD = 1.3$; $p = .009$). They also reported their attention being spread across fewer aspects of the plans, but no significant difference ($M = 5.1$, $SD = 1.3$ vs.\ $M = 5.4$, $SD = 1.1$; $p = .091$).
\end{itemize}

\subsubsection{CQ Quality Evaluations}
\label{sec:cq-evaluations}
Participants who encountered CQs rated the usefulness of the first and second CQs at means of 7.7 ($SD = 2.1$) and 8.0 ($SD = 1.8$) out of 10 across the 270 responses.
In the post-study items, they agreed that the CQs were relevant ($M = 5.6$, $SD = 1.0$, median 6 of 7, 91\% at or above 5).
Participants generally did not consider answering CQs posing more extra efforts (the effort item: $M = 2.8$, $SD = 1.5$, median 3, 17\%).
They moderately agreed that being asked CQs made them trust the final response more ($M = 5.3$, $SD = 1.0$, median 5, 81\%).

\section{Discussion}

Clarifying questions are commonly motivated by their ability to elicit information that helps an LLM better address users' needs. 
\textbf{Our findings suggest that their influence extends to how users subsequently engage with and rely on the resulting response.}
Participants who encountered CQs spent less time reviewing and evaluating responses and were more willing to use them unchanged, alongside higher perceived understanding and satisfaction. 
Within the CQ condition, violations concerning previously queried sub-decisions were also more likely to be retained than violations concerning unqueried sub-decisions.

These findings raise a broader question for proactive LLM interaction: how does an intervention that helps specify a task change users' subsequent evaluation of its outcome? 
We discuss the benefits and consequences of this shift, consider possible mechanisms and alternative explanations, and identify implications for supporting continued evaluation after successful clarification.    

\subsection{Proactive Clarification Reshapes What Happens After LLM Recommendation}

Our results first reinforce the benefits of proactive context elicitation.
Participants who encountered CQs felt better understood, were more satisfied with the assistant, and submitted fewer preference-based requests after receiving the response. 
This reduction suggests that CQ addressed preferences that would otherwise require subsequent revision. 
These findings are consistent with research showing that context-seeking can make LLM responses feel more relevant and tailored~\cite{sayres2026towards}.

However, these benefits accompanied changes in subsequent engagement and reliance. Participants who encountered CQs spent less time reviewing and evaluating responses and were more likely to accept both constraint-following and erroneous responses unchanged. 
Error reporting and retrospective awareness were also descriptively lower after answering CQs.
Their nonsignificant differences do not indicate an absence of effects, but suggest effects smaller than those observed for engagement and reliance, which would require a larger sample to estimate with sufficient precision.
Shorter review following CQs also occurred on constraint-following trials, suggesting that the time difference was not solely attributable to participants spending longer responding to errors. 
This pattern points to a broader change in how participants approached the assistant’s answer. 
CQs therefore functioned not only as an information-elicitation technique, but also as an intervention that shaped subsequent evaluation and reliance.

This finding has implications for the broader evaluation of proactive LLM interaction.
Prior work has examined whether proactive interventions are timely, relevant, and well received by users~\cite{kuo2026developer, sayres2026towards}. 
From a mixed-initiative perspective, however, system and user initiatives are interdependent rather than isolated~\cite{horvitz1999principles}: when an assistant takes initiative to clarify a task, it may also change whether users subsequently take initiative to inspect, question, or revise its response. 
Resolving ambiguity may reduce unnecessary follow-up by better addressing users’ needs, but it may also lead to side effects of reducing scrutiny of problems that remain. \textbf{Evaluations of proactive interaction should therefore examine how AI system initiative redistributes user initiative across the subsequent interaction, alongside its immediate effects on output quality and user experience.}

\subsection{Why Clarification May Change Subsequent Engagement and Reliance}

Two features of CQs may each lead to our observations in the study, and the two may add up. 
First, \textbf{answering a question and seeing the answer reflected in the response is a grounding act} (\autoref{sec:grounding}), which can make participants feel that the relevant sub-decisions were settled. Second, \textbf{the LLM response is personalized based on participants answers to CQs}, which can make participants feel the sub-decision fit their needs well.
The first feature is about the interaction process and the second is about the LLM outputs.
Deployed CQs with LLM systems carry both features at once. 
Our study treats CQ, a human-LLM interaction technique, as the study unit.
Thus, the study design did not aim to separate the effects of the two features, which warrants future research.
Below we discuss each explanation, and further consider user familiarity as another alternative explanation to the timing results.

\subsubsection{Clarification as Grounding}

Grounding establishes shared understanding sufficient for the current purpose~\cite{clark1991grounding}. 
Interactions around CQs may meet the grounding at the level of a sub-decision. 
By asking users to articulate a preference and visibly incorporating it into the response, LLM assistants provide evidence that the user input for that sub-decision has been understood and acted upon. 
The sub-decision may consequently feel sufficiently handled. 
We refer to the possibility as \emph{perceived task resolution}, meaning that an aspect of the task is treated as grounded and therefore as needing less further evaluation.

Perceived resolution may influence reliance by changing the perceived value of further verification. 
Prior work suggests that users allocate effort by weighing the benefits of checking AI advice against the effort required~\cite{vasconcelos2023explanations}. 
Once an assistant has asked about a sub-decision and incorporated the user's answer, revisiting it may appear redundant. 
This creates a risk, however, in the human-LLM interaction contexts.
As highlighted in our study, the elicited preference and the associated hard constraint can be distinct. 
An assistant could follow an elicited user preference for self-installation while recommending a component that arrived after the installation deadline.
A completed exchange about the user preference may therefore negatively impacts the evaluation of the LLM answers' correctness.

Several findings are consistent with the grounding account.
In the post-study survey, participants who encountered CQs felt that the assistant better understood their intentions, reported fewer criteria to keep track of, and perceived less need to double-check its responses. 
These self-reported perceptions accompanied shorter review and evaluation times. 
At the same time, general trust in the assistant did not differ between conditions. 
The contrast suggests a shift specific to the specific task rather than to their mental model of the LLM assistant. 
\textbf{Participants may not increase their trust to the assistant with CQs, but may have regarded the particular task as more fully specified and the resulting LLM response as requiring less additional work.}
The account also predicts that the effect should attach to the sub-decisions that were discussed.
In the w/ CQ condition of our study, the added acceptance of constraint-violating content stayed within the sub-decisions the assistant had asked about, even though review time and reported checking needs changed across the whole response.
Higher confidence and usefulness ratings for responses with adjacent violations is also complementary evidence.

The account leaves open where the drop in scrutiny occurred. Recognition and corrective action need not coincide, and some participants retained a violating paragraph despite retrospectively reporting that they had noticed the issue.
The gap between noticing and acting parallels metacognitive monitoring and control~\cite{tankelevitch2024metacognitive} and observations of users adopting erroneous LLM content despite hesitation~\cite{liu2026behavioral}.
Clarification may influence either process, although our retrospective measures cannot determine which.

\subsubsection{Clarification as Personalization}

The second possible explanation of CQs impacts starts from the content of the LLM response rather than the interaction process. 
Paragraphs in CQ-adjacent sub-decisions followed the participant's own choice, whereas the paragraph in the CQ-nonadjacent sub-decision followed a randomly selected option. 
The design kept the practical purpose of CQs, since elicited preferences should shape the LLM recommendation, and the drop in preference-based requests confirms the benefit. 
Preference fit predicts the same scope as grounding. 
When reading a LLM paragraph that follows user's just-elicited preference, users might feel the paragraph as a well-considered response to their needs and thus quickly skim without closer look.
Conversely, paragraphs that do not visibly reflect an elicited preference may lack the cue of responsiveness and therefore receive greater scrutiny.

The higher rate of error reporting for unasked sub-decisions fits either account, since a closer look finds more errors whatever caused it. 
The two explanations therefore make the same prediction for the adjacency contrast and for the comparison with the no-CQ condition, and our data do not separate them. 
They differ in emphasis. 
Personalization locates the effect in what the response says, and grounding locates it in what the interaction did.
Arguably, personalization alone may not predict the lower reported need to double-check or the smaller set of criteria participants felt they had to track, which concern the response as a whole.
We interpret our study results as the two acting together, with personalization accounting for much of where the added paragraph-level acceptance landed and perceived resolution accounting for the response-level drop in evaluative effort.

\subsubsection{Alternative Explanation}
Clarification may increase familiarity with the topics and options discussed. 
Participants had already considered two sub-decisions, potentially making the response easier and faster to process. 
This provides a plausible explanation for shorter review.
However, processing efficiency alone does not explain the accompanying reduction in perceived checking needs or greater acceptance of CQ-adjacent violations. 
Familiarity may also shape the focus of evaluation, encouraging users to confirm that the response reflects their selected option without revisiting every associated requirement. 
Thus, familiarity may contribute to the timing difference while interacting with a broader change in what participants consider necessary to check.

\subsection{Design Implications for Human-LLM Interaction}

Our findings motivate how systems might preserve the benefits of CQs while supporting continued evaluation when needed.

\paragraph{Provide user-side cues about the scope of resolution.}
Proactive systems could provide user interface cues about what a preceding exchange has established and what remains unresolved. 
These cues might indicate which aspects of a recommendation were personalized, which requirements were explicitly checked, and which conclusions depend on assumptions or uncertain information. 
The purpose of such cues would be to help users distinguish evidence that the assistant understood their input from evidence that the recommendation is well-established.
Simply marking an entire topic as ``covered'' could instead reinforce the impression that it needs no further attention. 

\paragraph{System-side verification where user clarification may reduce scrutiny}
Participants were least likely to flag violations in sub-decisions that the assistant had previously asked about, and the assistant can identify which parts of its response correspond to those clarification exchanges. 
LLM Systems could use this interaction history to apply additional checks when generating content that incorporates elicited preferences. 
Such targeted checking does not replace user oversight.
It treats user and system scrutiny as complementary resources by increasing system-side verification where the clarification exchange may have reduced users' own scrutiny.

\paragraph{Reactivate evaluation through cognitive forcing}
Cognitive forcing may help users re-engage with a response that appears sufficiently resolved. 
Prior work shows that interventions encouraging deliberation can reduce overreliance on AI advice~\cite{buccinca2021trust, ma2025towards}, and that reasoning-oriented questions can improve discernment~\cite{danry2023don}. 
In our context, such interventions could direct users back to consequential constraints or unresolved assumptions after preference elicitation. 
One example is a proactive post-answer question explicitly asking users to compare a proposed schedule against their requirements.
The purpose is to prompt substantive evaluation, rather than simply obtain another confirmation. 
Because these interventions also impose effort, they should target aspects where additional scrutiny is valuable and be evaluated for both corrective benefits and interaction burden.

\subsection{Limitations and Future Work}

Our behavioral measures and post-study ratings may not capture the subtler cognitive processes through which CQs shape evaluation.
They nevertheless allowed us to examine subsequent engagement without repeatedly disrupting participants interaction with LLMs.
Future work could combine unobtrusive attention measures with carefully timed assessments to distinguish faster comprehension, changes in evaluative focus, and a reduced perceived need to check. 
Larger samples would also help estimate the observed awareness differences more precisely.

Besides, our scenarios represented everyday decisions but did not carry the personal consequences of implementing the recommendations. 
The task scenarios used in our study enabled controlled comparisons across several decision contexts, while matching response structure and approximate length controls the confounding of presentation differences. 
These choices are dedicated for the controlled study but leave open whether users would allocate effort differently when facing personal stakes or substantially longer responses. 
Future studies should examine consequential decisions and vary response length and complexity to test the boundaries of the observed effects.

Lastly, we focused on constraint violations that could be identified by comparing the response with the scenario requirements, instead of including more potential LLM mistakes.
This provided a controlled setting for examining critical evaluation, particularly given the low detection of factual errors in our pilot. 
The findings may differ for factual hallucinations, or errors that directly contradict elicited preferences from CQs.
Likewise, fixing the number and format of CQs supported a consistent comparison but did not capture impacts of design factors of CQs. 
Future studies should vary error types and clarification strategies, including question timing and optionality.

\section{Conclusion}

Using questions to have user clarification of their preferences does not ensure that the resulting LLM recommendation is perfect.
Our findings highlight the importance of this distinction.
Users can feel better understood and become more willing to accept an answer even when it contains unresolved problems, particularly within the sub-decisions involved in clarification. 
The value of CQs therefore depends partly on what follows the exchange around them.
As LLM assistants become more proactive, their success should be assessed through both the immediate benefits of eliciting context and the subsequent engagement and reliance they encourage. 
Supporting users requires helping them communicate what they want while keeping remaining assumptions and constraints open to examination.

\bibliographystyle{ACM-Reference-Format}
\bibliography{references}

@inproceedings{ashktorab2019resilient,
  author =        {Ashktorab, Zahra and Jain, Mohit and Liao, Q Vera and
                   Weisz, Justin D},
  booktitle =     {Proceedings of the 2019 CHI conference on human
                   factors in computing systems},
  pages =         {1--12},
  title =         {Resilient chatbots: Repair strategy preferences for
                   conversational breakdowns},
  year =          {2019},
}

@inproceedings{bansal2021does,
  author =        {Bansal, Gagan and Wu, Tongshuang and Zhou, Joyce and
                   Fok, Raymond and Nushi, Besmira and Kamar, Ece and
                   Ribeiro, Marco Tulio and Weld, Daniel},
  booktitle =     {Proceedings of the 2021 CHI conference on human
                   factors in computing systems},
  pages =         {1--16},
  title =         {Does the whole exceed its parts? the effect of ai
                   explanations on complementary team performance},
  year =          {2021},
}

@inproceedings{benotti2021grounding,
  author =        {Benotti, Luciana and Blackburn, Patrick},
  booktitle =     {Proceedings of the 16th conference of the European
                   chapter of the association for computational
                   linguistics: Main volume},
  pages =         {515--531},
  title =         {Grounding as a collaborative process},
  year =          {2021},
}

@inproceedings{bondarenko2022user,
  author =        {Bondarenko, Alexander and Shirshakova, Ekaterina and
                   Hagen, Matthias},
  booktitle =     {Proceedings of the 2022 Conference on Human
                   Information Interaction and Retrieval},
  pages =         {254--258},
  title =         {A user study on clarifying comparative questions},
  year =          {2022},
}

@article{buccinca2021trust,
  author =        {Bu{\c{c}}inca, Zana and Malaya, Maja Barbara and
                   Gajos, Krzysztof Z},
  journal =       {Proceedings of the ACM on Human-computer Interaction},
  number =        {CSCW1},
  pages =         {1--21},
  publisher =     {ACM New York, NY, USA},
  title =         {To trust or to think: cognitive forcing functions can
                   reduce overreliance on AI in AI-assisted
                   decision-making},
  volume =        {5},
  year =          {2021},
}

@techreport{chatterji2025people,
  author =        {Chatterji, Aaron and Cunningham, Thomas and
                   Deming, David J and Hitzig, Zoe and Ong, Christopher and
                   Shan, Carl Yan and Wadman, Kevin},
  institution =   {National Bureau of Economic Research},
  title =         {How people use chatgpt},
  year =          {2025},
}

@inproceedings{chen2025dango,
  author =        {Chen, Wei-Hao and Tong, Weixi and Case, Amanda and
                   Zhang, Tianyi},
  booktitle =     {Proceedings of the 2025 CHI Conference on Human
                   Factors in Computing Systems},
  pages =         {1--28},
  title =         {Dango: a mixed-initiative data wrangling system using
                   large language model},
  year =          {2025},
}

@article{clark1991grounding,
  author =        {Clark, Herbert H and Brennan, Susan E},
  publisher =     {American Psychological Association},
  title =         {Grounding in communication.},
  year =          {1991},
}

@inproceedings{danry2023don,
  author =        {Danry, Valdemar and Pataranutaporn, Pat and
                   Mao, Yaoli and Maes, Pattie},
  booktitle =     {Proceedings of the 2023 CHI conference on human
                   factors in computing systems},
  pages =         {1--13},
  title =         {Don’t just tell me, ask me: Ai systems that
                   intelligently frame explanations as questions improve
                   human logical discernment accuracy over causal ai
                   explanations},
  year =          {2023},
}

@incollection{hart1988development,
  author =        {Hart, Sandra G and Staveland, Lowell E},
  booktitle =     {Advances in psychology},
  pages =         {139--183},
  publisher =     {Elsevier},
  title =         {Development of NASA-TLX (Task Load Index): Results of
                   empirical and theoretical research},
  volume =        {52},
  year =          {1988},
}

@inproceedings{horvitz1999principles,
  author =        {Horvitz, Eric},
  booktitle =     {Proceedings of the SIGCHI conference on Human Factors
                   in Computing Systems},
  pages =         {159--166},
  title =         {Principles of mixed-initiative user interfaces},
  year =          {1999},
}

@article{hu2024designing,
  author =        {Hu, Jiaxiong and Guo, Jingya and Tang, Ningjing and
                   Ma, Xiaojuan and Yao, Yuan and Yang, Changyuan and
                   Xu, Yingqing},
  journal =       {Proceedings of the ACM on Human-Computer Interaction},
  number =        {CSCW1},
  pages =         {1--30},
  publisher =     {ACM New York, NY, USA},
  title =         {Designing the conversational agent: asking follow-up
                   questions for information elicitation},
  volume =        {8},
  year =          {2024},
}

@article{huang2025survey,
  author =        {Huang, Lei and Yu, Weijiang and Ma, Weitao and
                   Zhong, Weihong and Feng, Zhangyin and Wang, Haotian and
                   Chen, Qianglong and Peng, Weihua and Feng, Xiaocheng and
                   Qin, Bing and others},
  journal =       {ACM transactions on information systems},
  number =        {2},
  pages =         {1--55},
  publisher =     {ACM New York, NY},
  title =         {A survey on hallucination in large language models:
                   Principles, taxonomy, challenges, and open questions},
  volume =        {43},
  year =          {2025},
}

@inproceedings{jansen2025user,
  author =        {Jansen, Bernard J and Guan, Kathleen W and
                   Salminen, Joni and Aldous, Kholoud Khalil and
                   Jung, Soon-Gyo},
  booktitle =     {Proceedings of the 2025 CHI conference on human
                   factors in computing systems},
  pages =         {1--19},
  title =         {What is user engagement?: A systematic review of 241
                   research articles in human-computer interaction and
                   beyond},
  year =          {2025},
}

@article{jian2000foundations,
  author =        {Jian, Jiun-Yin and Bisantz, Ann M and Drury, Colin G},
  journal =       {International journal of cognitive ergonomics},
  number =        {1},
  pages =         {53--71},
  publisher =     {Taylor \& Francis},
  title =         {Foundations for an empirically determined scale of
                   trust in automated systems},
  volume =        {4},
  year =          {2000},
}

@inproceedings{jung2026ask,
  author =        {Jung, Gyuyeon and Ho, Hua Yun and Quartson, Pamela and
                   Cheshire, Coye},
  booktitle =     {Proceedings of the 2026 Conference on Creativity and
                   Cognition},
  pages =         {1503--1509},
  title =         {Ask Before or After? Clarification Timing and User
                   Experience in Generative AI},
  year =          {2026},
}

@inproceedings{kim2024m,
  author =        {Kim, Sunnie SY and Liao, Q Vera and
                   Vorvoreanu, Mihaela and Ballard, Stephanie and
                   Vaughan, Jennifer Wortman},
  booktitle =     {Proceedings of the 2024 ACM conference on fairness,
                   accountability, and transparency},
  pages =         {822--835},
  title =         {" I'm Not Sure, But...": Examining the Impact of
                   Large Language Models' Uncertainty Expression on User
                   Reliance and Trust},
  year =          {2024},
}

@inproceedings{kim2025fostering,
  author =        {Kim, Sunnie SY and Vaughan, Jennifer Wortman and
                   Liao, Q Vera and Lombrozo, Tania and
                   Russakovsky, Olga},
  booktitle =     {Proceedings of the 2025 CHI Conference on Human
                   Factors in Computing Systems},
  pages =         {1--19},
  title =         {Fostering appropriate reliance on large language
                   models: The role of explanations, sources, and
                   inconsistencies},
  year =          {2025},
}

@inproceedings{kim2024understanding,
  author =        {Kim, Yoonsu and Lee, Jueon and Kim, Seoyoung and
                   Park, Jaehyuk and Kim, Juho},
  booktitle =     {Proceedings of the 29th International Conference on
                   Intelligent User Interfaces},
  pages =         {385--404},
  title =         {Understanding users’ dissatisfaction with chatgpt
                   responses: Types, resolving tactics, and the effect
                   of knowledge level},
  year =          {2024},
}

@inproceedings{kuo2026developer,
  author =        {Kuo, Nadine and Sergeyuk, Agnia and Chen, Valerie and
                   Izadi, Maliheh},
  booktitle =     {Proceedings of the 31st International Conference on
                   Intelligent User Interfaces},
  pages =         {349--362},
  title =         {Developer interaction patterns with proactive AI: A
                   five-day field study},
  year =          {2026},
}

@inproceedings{li2025eliciting,
  author =        {Li, Belinda and Tamkin, Alex and Goodman, Noah and
                   Andreas, Jacob},
  booktitle =     {International Conference on Learning Representations},
  pages =         {80984--81013},
  title =         {Eliciting human preferences with language models},
  volume =        {2025},
  year =          {2025},
}

@inproceedings{liu2026behavioral,
  author =        {Liu, Chang and Zhou, Qinyi and Shen, Xinjie and
                   Bruce Liu, Xingyu and Wu, Tongshuang and
                   Chen, Xiang'Anthony},
  booktitle =     {Proceedings of the 2026 CHI Conference on Human
                   Factors in Computing Systems},
  pages =         {1--23},
  title =         {Behavioral indicators of overreliance during
                   interaction with conversational language models},
  year =          {2026},
}

@inproceedings{ma2025towards,
  author =        {Ma, Shuai and Chen, Qiaoyi and Wang, Xinru and
                   Zheng, Chengbo and Peng, Zhenhui and Yin, Ming and
                   Ma, Xiaojuan},
  booktitle =     {Proceedings of the 2025 CHI Conference on Human
                   Factors in Computing Systems},
  pages =         {1--23},
  title =         {Towards human-ai deliberation: Design and evaluation
                   of llm-empowered deliberative ai for ai-assisted
                   decision-making},
  year =          {2025},
}

@inproceedings{ma2023should,
  author =        {Ma, Shuai and Lei, Ying and Wang, Xinru and
                   Zheng, Chengbo and Shi, Chuhan and Yin, Ming and
                   Ma, Xiaojuan},
  booktitle =     {Proceedings of the 2023 CHI Conference on Human
                   Factors in Computing Systems},
  pages =         {1--19},
  title =         {Who should i trust: Ai or myself? leveraging human
                   and ai correctness likelihood to promote appropriate
                   trust in ai-assisted decision-making},
  year =          {2023},
}

@article{mccrudden2007relevance,
  author =        {McCrudden, Matthew T and Schraw, Gregory},
  journal =       {Educational psychology review},
  number =        {2},
  pages =         {113--139},
  publisher =     {Springer},
  title =         {Relevance and goal-focusing in text processing},
  volume =        {19},
  year =          {2007},
}

@article{mcgrath2025measuring,
  author =        {McGrath, Melanie J and Lack, Oliver and Tisch, James and
                   Duenser, Andreas},
  journal =       {Frontiers in Artificial Intelligence},
  pages =         {1582880},
  publisher =     {Frontiers Media SA},
  title =         {Measuring trust in artificial intelligence:
                   validation of an established scale and its short
                   form},
  volume =        {8},
  year =          {2025},
}

@inproceedings{niu2024ragtruth,
  author =        {Niu, Cheng and Wu, Yuanhao and Zhu, Juno and
                   Xu, Siliang and Shum, Kashun and Zhong, Randy and
                   Song, Juntong and Zhang, Tong},
  booktitle =     {Proceedings of the 62nd Annual Meeting of the
                   Association for Computational Linguistics (Volume 1:
                   Long Papers)},
  pages =         {10862--10878},
  title =         {Ragtruth: A hallucination corpus for developing
                   trustworthy retrieval-augmented language models},
  year =          {2024},
}

@article{norman2010likert,
  author =        {Norman, Geoff},
  journal =       {Advances in health sciences education},
  number =        {5},
  pages =         {625--632},
  publisher =     {Springer},
  title =         {Likert scales, levels of measurement and the
                   “laws” of statistics},
  volume =        {15},
  year =          {2010},
}

@inproceedings{salimzadeh2024dealing,
  author =        {Salimzadeh, Sara and He, Gaole and Gadiraju, Ujwal},
  booktitle =     {Proceedings of the 2024 CHI Conference on Human
                   Factors in Computing Systems},
  pages =         {1--17},
  title =         {Dealing with uncertainty: Understanding the impact of
                   prognostic versus diagnostic tasks on trust and
                   reliance in human-AI decision making},
  year =          {2024},
}

@inproceedings{sayres2026towards,
  author =        {Sayres, Rory and Hao, Yuexing and Ward, Abbi and
                   Wang, Amy and Freeman, Beverly and Zhan, Serena and
                   Ardila, Diego and Li, Jimmy and Lee, I-Ching and
                   Iurchenko, Anna and others},
  booktitle =     {Proceedings of the 2026 CHI Conference on Human
                   Factors in Computing Systems},
  pages =         {1--28},
  title =         {Towards Better Health Conversations: The Benefits of
                   Context-seeking},
  year =          {2026},
}

@inproceedings{schemmer2023appropriate,
  author =        {Schemmer, Max and Kuehl, Niklas and Benz, Carina and
                   Bartos, Andrea and Satzger, Gerhard},
  booktitle =     {Proceedings of the 28th international conference on
                   intelligent user interfaces},
  pages =         {410--422},
  title =         {Appropriate reliance on AI advice: Conceptualization
                   and the effect of explanations},
  year =          {2023},
}

@inproceedings{shaikh2024grounding,
  author =        {Shaikh, Omar and Gligori{\'c}, Kristina and
                   Khetan, Ashna and Gerstgrasser, Matthias and
                   Yang, Diyi and Jurafsky, Dan},
  booktitle =     {Proceedings of the 2024 Conference of the North
                   American Chapter of the Association for Computational
                   Linguistics: Human Language Technologies (Volume 1:
                   Long Papers)},
  pages =         {6279--6296},
  title =         {Grounding gaps in language model generations},
  year =          {2024},
}

@inproceedings{sharma2026feedback,
  author =        {Sharma, Nikhil and Zhang, Zheng and Lee, Daniel and
                   Krishnan, Namita and Ren, Guang-Jie and Xiao, Ziang and
                   Li, Yunyao},
  booktitle =     {Proceedings of the 2026 CHI Conference on Human
                   Factors in Computing Systems},
  pages =         {1--22},
  title =         {Feedback by design: Understanding and overcoming user
                   feedback barriers in conversational agents},
  year =          {2026},
}

@inproceedings{subramonyam2024bridging,
  author =        {Subramonyam, Hari and Pea, Roy and
                   Pondoc, Christopher and Agrawala, Maneesh and
                   Seifert, Colleen},
  booktitle =     {Proceedings of the 2024 CHI Conference on Human
                   Factors in Computing Systems},
  pages =         {1--19},
  title =         {Bridging the gulf of envisioning: Cognitive
                   challenges in prompt based interactions with LLMs},
  year =          {2024},
}

@inproceedings{tankelevitch2024metacognitive,
  author =        {Tankelevitch, Lev and Kewenig, Viktor and
                   Simkute, Auste and Scott, Ava Elizabeth and
                   Sarkar, Advait and Sellen, Abigail and Rintel, Sean},
  booktitle =     {Proceedings of the 2024 CHI conference on human
                   factors in computing systems},
  pages =         {1--24},
  title =         {The metacognitive demands and opportunities of
                   generative AI},
  year =          {2024},
}

@inproceedings{tix2025follow,
  author =        {Tix, Bernadette J},
  booktitle =     {International Conference on Human-Computer
                   Interaction},
  organization =  {Springer},
  pages =         {288--302},
  title =         {Follow-up questions improve documents generated by
                   large language models},
  year =          {2025},
}

@article{tou2025shoppingcomp,
  author =        {Tou, Huaixiao and Zeng, Ying and Li, Yuemeng and
                   Ma, Cong and Li, Muzhi and Li, Minghao and
                   Yuan, Weijie and Zhang, He and Jia, Kai},
  journal =       {arXiv preprint arXiv:2511.22978},
  title =         {ShoppingComp: Are LLMs Really Ready for Your Shopping
                   Cart?},
  year =          {2025},
}

@article{vasconcelos2023explanations,
  author =        {Vasconcelos, Helena and J{\"o}rke, Matthew and
                   Grunde-McLaughlin, Madeleine and Gerstenberg, Tobias and
                   Bernstein, Michael S and Krishna, Ranjay},
  journal =       {Proceedings of the ACM on human-computer interaction},
  number =        {CSCW1},
  pages =         {1--38},
  publisher =     {ACM New York, NY, USA},
  title =         {Explanations can reduce overreliance on ai systems
                   during decision-making},
  volume =        {7},
  year =          {2023},
}

@inproceedings{wen2026adaptive,
  author =        {Wen, Xinyi and Hegemann, Lena and Jin, Xiaofu and
                   Ma, Shuai and Oulasvirta, Antti},
  booktitle =     {Proceedings of the 31st International Conference on
                   Intelligent User Interfaces},
  pages =         {730--754},
  title =         {Adaptive Prompt Elicitation for Text-to-Image
                   Generation},
  year =          {2026},
}

@article{xie2024travelplanner,
  author =        {Xie, Jian and Zhang, Kai and Chen, Jiangjie and
                   Zhu, Tinghui and Lou, Renze and Tian, Yuandong and
                   Xiao, Yanghua and Su, Yu},
  journal =       {arXiv preprint arXiv:2402.01622},
  title =         {Travelplanner: A benchmark for real-world planning
                   with language agents},
  year =          {2024},
}

@article{yin2024ai,
  author =        {Yin, Yidan and Jia, Nan and Wakslak, Cheryl J},
  journal =       {Proceedings of the National Academy of Sciences},
  number =        {14},
  pages =         {e2319112121},
  publisher =     {National Academy of Sciences},
  title =         {AI can help people feel heard, but an AI label
                   diminishes this impact},
  volume =        {121},
  year =          {2024},
}

@inproceedings{zamani2020generating,
  author =        {Zamani, Hamed and Dumais, Susan and Craswell, Nick and
                   Bennett, Paul and Lueck, Gord},
  booktitle =     {Proceedings of the web conference 2020},
  pages =         {418--428},
  title =         {Generating clarifying questions for information
                   retrieval},
  year =          {2020},
}

@inproceedings{zamfirescu2023johnny,
  author =        {Zamfirescu-Pereira, J Diego and Wong, Richmond Y and
                   Hartmann, Bjoern and Yang, Qian},
  booktitle =     {Proceedings of the 2023 CHI conference on human
                   factors in computing systems},
  pages =         {1--21},
  title =         {Why Johnny can’t prompt: how non-AI experts try
                   (and fail) to design LLM prompts},
  year =          {2023},
}

@inproceedings{zhang2025clarify,
  author =        {Zhang, Michael JQ and Choi, Eunsol},
  booktitle =     {Findings of the Association for Computational
                   Linguistics: NAACL 2025},
  pages =         {5541--5558},
  title =         {Clarify when necessary: Resolving ambiguity through
                   interaction with lms},
  year =          {2025},
}

@inproceedings{zhang2024clamber,
  author =        {Zhang, Tong and Qin, Peixin and Deng, Yang and
                   Huang, Chen and Lei, Wenqiang and Liu, Junhong and
                   Jin, Dingnan and Liang, Hongru and Chua, Tat-Seng},
  booktitle =     {Proceedings of the 62nd Annual Meeting of the
                   Association for Computational Linguistics (Volume 1:
                   Long Papers)},
  pages =         {10746--10766},
  title =         {CLAMBER: A benchmark of identifying and clarifying
                   ambiguous information needs in large language models},
  year =          {2024},
}

@inproceedings{zheng2024disciplink,
  author =        {Zheng, Chengbo and Zhang, Yuanhao and Huang, Zeyu and
                   Shi, Chuhan and Xu, Minrui and Ma, Xiaojuan},
  booktitle =     {Proceedings of the 37th Annual ACM Symposium on User
                   Interface Software and Technology},
  pages =         {1--20},
  title =         {Disciplink: Unfolding interdisciplinary information
                   seeking process via human-ai co-exploration},
  year =          {2024},
}

@article{zou2023users,
  author =        {Zou, Jie and Aliannejadi, Mohammad and
                   Kanoulas, Evangelos and Pera, Maria Soledad and
                   Liu, Yiqun},
  journal =       {ACM transactions on information systems},
  number =        {1},
  pages =         {1--25},
  publisher =     {ACM New York, NY},
  title =         {Users meet clarifying questions: Toward a better
                   understanding of user interactions for search
                   clarification},
  volume =        {41},
  year =          {2023},
}
\appendix

\section{Counterbalancing Schedules}

\autoref{tab:schedules} lists the 18 schedules.

\label{app:schedules}
\begin{table*}[t]
  \caption{The 18 schedules for counterbalancing. Each schedule fixes three scenarios in order, the sub-decision that carries the constraint violation (-- for the correct scenario), and the two sub-decisions the assistant asks about. A marks a CQ-adjacent violation and N a CQ-nonadjacent one. Scenarios: APT apartment hunting, CAR car purchase, DIN group dinner, EAR earbuds shopping, LCK smart lock, NYC New York trip. The w/o CQ condition uses the same schedules with the questions omitted.}
  \label{tab:schedules}
  \footnotesize
  \setlength{\tabcolsep}{3pt}
  \begin{tabular}{r lll lll lll}
    \toprule
    & \multicolumn{3}{c}{First scenario} & \multicolumn{3}{c}{Second scenario} & \multicolumn{3}{c}{Third scenario} \\
    \cmidrule(lr){2-4} \cmidrule(lr){5-7} \cmidrule(lr){8-10}
    \# & Sc. & Violation & Asked about & Sc. & Violation & Asked about & Sc. & Violation & Asked about \\
    \midrule
    1 & CAR & dealer (N) & car, handover & DIN & -- & seating, bill & APT & apartment (A) & apartment, lease \\
    2 & DIN & bill (N) & restaurant, seating & EAR & -- & earbuds, charging & APT & lease (A) & lease, viewing \\
    3 & EAR & earbuds (N) & seller, charging & APT & viewing (A) & apartment, viewing & CAR & -- & car, dealer \\
    4 & DIN & seating (N) & restaurant, bill & CAR & car (A) & car, handover & EAR & -- & seller, charging \\
    5 & CAR & dealer (A) & car, dealer & EAR & charging (N) & earbuds, seller & LCK & -- & lock, access \\
    6 & DIN & -- & restaurant, seating & LCK & lock (N) & install, access & CAR & handover (A) & dealer, handover \\
    7 & LCK & -- & install, access & DIN & restaurant (A) & restaurant, seating & EAR & seller (N) & earbuds, charging \\
    8 & NYC & -- & hotel, afternoon & LCK & access (N) & lock, install & DIN & seating (A) & seating, bill \\
    9 & EAR & -- & earbuds, seller & DIN & bill (A) & restaurant, bill & NYC & hotel (N) & lunch, afternoon \\
    10 & EAR & earbuds (A) & earbuds, charging & NYC & -- & lunch, afternoon & LCK & install (N) & lock, access \\
    11 & APT & -- & apartment, viewing & EAR & seller (A) & earbuds, seller & NYC & afternoon (N) & hotel, lunch \\
    12 & APT & apartment (N) & lease, viewing & LCK & -- & lock, install & EAR & charging (A) & seller, charging \\
    13 & APT & -- & lease, viewing & NYC & lunch (N) & hotel, afternoon & LCK & lock (A) & lock, install \\
    14 & LCK & install (A) & install, access & APT & viewing (N) & apartment, lease & CAR & -- & car, handover \\
    15 & LCK & access (A) & lock, access & CAR & car (N) & dealer, handover & NYC & -- & hotel, lunch \\
    16 & NYC & hotel (A) & hotel, afternoon & CAR & -- & dealer, handover & APT & lease (N) & apartment, viewing \\
    17 & CAR & handover (N) & car, dealer & NYC & lunch (A) & hotel, lunch & DIN & -- & restaurant, bill \\
    18 & NYC & afternoon (A) & lunch, afternoon & APT & -- & apartment, lease & DIN & restaurant (N) & seating, bill \\
    \bottomrule
  \end{tabular}
\end{table*}

\section{Scenario Materials}
\label{app:materials}
\providecommand{\viol}[1]{\textcolor{red}{#1}}

Tables~\ref{tab:materials-a} and~\ref{tab:materials-b} list the 18 sub-decisions.
Each row gives the CQs with its two options, the hard requirement, and excerpts of the two paragraphs the assistant could give for one of the options.
The option in bold is the one the excerpts follow, and the sentence in bold is the one that carries it.
The sentence in red is the one that violates the constraints.
The paragraphs for the other option are built the similar way.

\clearpage
\begingroup
\footnotesize
\setlength{\tabcolsep}{4pt}
\renewcommand{\arraystretch}{1.1}
\begin{xltabular}{\textwidth}{@{} >{\raggedright\arraybackslash}p{0.5in} >{\hsize=1.1\hsize\raggedright\arraybackslash}X >{\hsize=0.7\hsize\raggedright\arraybackslash}X >{\hsize=1.1\hsize\raggedright\arraybackslash}X >{\hsize=1.1\hsize\raggedright\arraybackslash}X @{}}
\caption{Materials for the first three scenarios.}\label{tab:materials-a}\\
\toprule
    Sub-decision & Clarifying question and its options & Hard requirement & Correct paragraph & Violating paragraph \\
    \midrule
\endfirsthead
\multicolumn{5}{c}{\normalsize Table~\thetable. Materials for the first three scenarios (continued).}\\[8pt]
\toprule
    Sub-decision & Clarifying question and its options & Hard requirement & Correct paragraph & Violating paragraph \\
    \midrule
\endhead
\midrule
\multicolumn{5}{r}{\emph{Continued on next page}}\\
\endfoot
\bottomrule
\endlastfoot
    \multicolumn{5}{@{}l@{}}{\textbf{Earbuds shopping.} Choose replacement earbuds for daily cycling.} \\*
    \cmidrule(l{0pt}r{0pt}){1-5}
    \noalign{\nobreak}
    Earbuds & \emph{What matters more in the earbuds?} \newline Option 1: \textbf{Nothing that can fall off --- but something clamped to your head the whole ride} \newline Option 2: Buds you can pocket --- but either one can work loose on a rough road & The earbuds must let outside sound through. & Shokz OpenRun Pro 2 --- \$180. They rest on your cheekbones and leave both ear canals open, so outside sound comes straight through. \textbf{The titanium band goes around the back of your head, so there is nothing to drop in the road.} & Soundcore Sport X20 --- \$80. \viol{The sealed tips and active noise cancelling block outside traffic sound.} \textbf{The hooks rotate over each ear, so there is nothing to drop in the road.} \\
    \addlinespace[3pt]
    Seller & \emph{Where do you want to buy them?} \newline Option 1: \textbf{Free shipping, wait for the box --- but no earbuds for the next few rides} \newline Option 2: In your hands today --- but you pay extra for the speed & The seller must accept opened-earbud returns for at least 30 days after delivery. & \textbf{Buy them from Amazon, where standard shipping is \$0, so you pay for the earbuds and nothing on top.} \ldots{} Amazon's 30-day return window covers opened earbuds after they have been tried. & \textbf{Buy them from Best Buy, where standard shipping is \$0, so you pay for the earbuds and nothing on top.} \ldots{} \viol{Opened earbuds have a 15-day window after delivery.} \\
    \addlinespace[3pt]
    Charging & \emph{What's your preference on charging?} \newline Option 1: \textbf{Carry a top-up with you --- but one more thing to remember and lose} \newline Option 2: Just a cable at home --- but a flat pair means a quiet ride & Whatever charges them must be USB-C. & For charging, get the Anker Nano Power Bank at \$30. \ldots{} The connector that folds out of it is USB-C. \textbf{It drops in a pocket and charges them without an outlet, so you are not stuck with a dead pair halfway home.} & For charging, get the Anker PowerCore 5000 at \$20. \viol{It charges over micro-USB and only has a USB-A output.} \ldots{} \textbf{It drops in a pocket and charges them without an outlet, so you are not stuck with a dead pair halfway home.} \\
    \addlinespace[3pt]
    \midrule
    \multicolumn{5}{@{}l@{}}{\textbf{Smart lock.} Choose a lock and arrange its installation and guest access.} \\*
    \cmidrule(l{0pt}r{0pt}){1-5}
    \noalign{\nobreak}
    Lock & \emph{What matters more in the lock?} \newline Option 1: \textbf{A new lock on the door --- but your current keys stop working} \newline Option 2: Keep the deadbolt you have --- but a second box on the inside & The quoted lock setup must include a keypad for code entry at the door. & For the lock, I'd choose the Yale Assure Lock 2 Touchscreen, \$200. \textbf{It replaces the deadbolt you have and leaves nothing extra on the inside of the door.} The keypad is on the face, so anyone at the door can let themselves in with a code. & For the lock, I'd choose the Level Lock, \$250. \textbf{It replaces the deadbolt you have and leaves nothing extra on the inside of the door.} \viol{A compatible keypad is sold separately, but it is not included in this quote; this package opens by app or key.} \\
    \addlinespace[3pt]
    Install & \emph{What's your preference for the fitting?} \newline Option 1: \textbf{Put it on yourself --- but your problem if the door sticks} \newline Option 2: Have someone fit it --- but a visit charge on top & The lock must be installed within one week. & \textbf{For installation, I'd choose do it yourself, \$0.} \ldots{} Installation can be completed within five days. & \textbf{For installation, I'd choose do it yourself, \$0.} \ldots{} \viol{The mounting bracket takes two weeks to arrive.} \\
    \addlinespace[3pt]
    Access & \emph{How should access work around the neighbor's daily visits?} \newline Option 1: \textbf{Keep access active --- but the code stays active all day between your neighbor's visits} \newline Option 2: Approve each day's visit --- but your neighbor waits outside if you haven't approved that day's visit & An entry method must be available for all eleven daily cat-sitting visits, the 12th to the 22nd. & For your neighbor's access, I'd keep access active the whole time, \$40. It is an 11-day access schedule starting on the 12th. \ldots{} \textbf{You do not need to activate access before each visit, but access stays active between visits.} & For your neighbor's access, I'd choose the default access schedule, \$0. It is \viol{a 7-day access schedule starting on the 12th.} \ldots{} \textbf{You do not need to activate access before each visit, but access stays active between visits.} \\
    \addlinespace[3pt]
    \midrule
    \multicolumn{5}{@{}l@{}}{\textbf{New York trip.} Plan a Saturday and one night for two.} \\*
    \cmidrule(l{0pt}r{0pt}){1-5}
    \noalign{\nobreak}
    Hotel & \emph{What's your preference for the hotel?} \newline Option 1: \textbf{Quieter, more character --- but a longer walk with the bags} \newline Option 2: Central, livelier street --- but traffic noise at the front & The room, with mandatory hotel fees, must stay under \$360 before tax. & Room: the Library Hotel. The base rate is \$315 for the night. \ldots{} Every room has a mandatory destination fee of \$25 a night. \textbf{It sits on a quiet side street with some character.} & Room: the Refinery Hotel. The base rate is \$339 for the night. \ldots{} \viol{Every room has a mandatory destination fee of \$45 a night.} \textbf{It sits on a quiet side street with some character.} \\
    \addlinespace[3pt]
    Lunch & \emph{What's your preference for lunch?} \newline Option 1: \textbf{Sit down, take your time --- but pricier than eating off a counter} \newline Option 2: Something quick off the counter --- but nowhere to sit at an unhurried pace & Saturday lunch must be available around 12:30. & For lunch, Cho Dang Gol, eight minutes from Bryant Park and \$26 per person. It serves lunch at 12:30 on Saturday. \textbf{It is a room with proper tables and a menu you read sitting down.} & For lunch, Keens Steakhouse, seven minutes from Bryant Park and \$48 per person. \viol{Saturday's earliest service is 13:45.} \textbf{It is an old chophouse, a room with proper tables and a menu you read sitting down.} \\
    \addlinespace[3pt]
    Afternoon & \emph{What's your preference for Saturday afternoon?} \newline Option 1: \textbf{One place, done at an unhurried pace --- but the rest of the city goes unseen} \newline Option 2: Several stops, easy pace --- but only a quick look at each & Nothing for the afternoon may need booking and paying for ahead. & \textbf{The afternoon: the Museum of Modern Art, done at an unhurried pace.} \ldots{} You buy at the door on the day. & \textbf{The afternoon: the Tenement Museum, done at an unhurried pace.} \ldots{} \viol{Admission is by a guided tour booked and paid for ahead.} \\
    \addlinespace[3pt]
\end{xltabular}
\endgroup

\clearpage
\begingroup
\footnotesize
\setlength{\tabcolsep}{4pt}
\renewcommand{\arraystretch}{1.1}
\begin{xltabular}{\textwidth}{@{} >{\raggedright\arraybackslash}p{0.5in} >{\hsize=1.1\hsize\raggedright\arraybackslash}X >{\hsize=0.7\hsize\raggedright\arraybackslash}X >{\hsize=1.1\hsize\raggedright\arraybackslash}X >{\hsize=1.1\hsize\raggedright\arraybackslash}X @{}}
\caption{Materials for the other three scenarios.}\label{tab:materials-b}\\
\toprule
    Sub-decision & Clarifying question and its options & Hard requirement & Correct paragraph & Violating paragraph \\
    \midrule
\endfirsthead
\multicolumn{5}{c}{\normalsize Table~\thetable. Materials for the other three scenarios (continued).}\\[8pt]
\toprule
    Sub-decision & Clarifying question and its options & Hard requirement & Correct paragraph & Violating paragraph \\
    \midrule
\endhead
\midrule
\multicolumn{5}{r}{\emph{Continued on next page}}\\
\endfoot
\bottomrule
\endlastfoot
    \multicolumn{5}{@{}l@{}}{\textbf{Apartment hunting.} Choose an apartment in Chicago.} \\*
    \cmidrule(l{0pt}r{0pt}){1-5}
    \noalign{\nobreak}
    Apartment & \emph{What's your preference for the apartment?} \newline Option 1: \textbf{More room for the money --- but a longer ride home after midnight} \newline Option 2: Smaller, a shorter ride --- but living on top of your own things & Rent plus required heat and hot water must not exceed \$1,550 a month. & I'd go with Rogers Park. \textbf{It is \$1,395 a month for 780 square feet, so there is room for a desk outside the bedroom.} Heat and hot water are another \$95 a month. & I'd go with Edgewater. \textbf{It is \$1,490 a month for 820 square feet, so there is room for a desk outside the bedroom.} \viol{Heat and hot water are another \$105 a month.} \\
    \addlinespace[3pt]
    Lease & \emph{What's your preference on the start date?} \newline Option 1: \textbf{Start as soon as possible --- but little time to sort the move out} \newline Option 2: Start later, get money off your move-in costs --- but weeks more in the old place & The refundable security deposit must not exceed one month's rent. & Start date: take the first-of-the-month start. \textbf{Keys in ten days, so you are out of the old place weeks sooner.} You pay one month's rent as the deposit. & Start date: take the immediate start. \textbf{Keys in three days, so you are out of the old place weeks sooner.} \viol{The deposit due at signing is two months' rent.} \\
    \addlinespace[3pt]
    Viewing & \emph{Would you rather see it on a weeknight or Saturday morning?} \newline Option 1: \textbf{A weeknight slot --- but seeing the place after dark} \newline Option 2: Saturday morning open house --- but going round it with everyone else & The viewing must happen within the next seven days. & For the showing, take the weeknight slot with the listing agent, two days from now. \ldots{} \textbf{You see the block on an ordinary weeknight.} & For the showing, take the agent's next weeknight opening, \viol{nine days from now.} \ldots{} \textbf{You see the block on an ordinary weeknight.} \\
    \addlinespace[3pt]
    \midrule
    \multicolumn{5}{@{}l@{}}{\textbf{Car purchase.} Choose a first car and arrange handover within 30 days.} \\*
    \cmidrule(l{0pt}r{0pt}){1-5}
    \noalign{\nobreak}
    Car & \emph{Would you rather have an EV or a hybrid with gas backup?} \newline Option 1: \textbf{EV for electric driving --- but charging stops on longer trips} \newline Option 2: Hybrid with gas backup --- but still buying gasoline & The advertised running-cost estimate must not assume charging at home. & For the car, I'd choose the Hyundai Ioniq 5 --- \$35,400. \ldots{} The broker's running-cost estimate uses public-charging prices, so it does not assume a home plug. \textbf{You do not buy gasoline for the town miles.} & For the car, I'd choose the Fiat 500e --- \$33,900. \ldots{} \viol{The broker's quoted running cost assumes overnight home charging.} \textbf{You do not buy gasoline for the town miles.} \\
    \addlinespace[3pt]
    Dealer & \emph{What's your preference on how you buy it?} \newline Option 1: \textbf{Pick the one you actually want --- but signing for a car you have not seen} \newline Option 2: Take one off the lot --- but paying for extras you did not pick & The seller must handle the title and plates. & How to buy it: order it in through Northgate Auto Center. \textbf{You pick the trim and color, and Northgate brings that car in.} \ldots{} The store files the title and puts the plates on. & How to buy it: the matching car at Fairlane Motors, two states away. \textbf{You pick the trim and color, and Fairlane already has that car.} \ldots{} \viol{Title and registration service is not included; buyer filing follows delivery.} \\
    \addlinespace[3pt]
    Handover & \emph{What's your preference for the handover?} \newline Option 1: \textbf{Have it brought to you --- but a fee for a trip you could make} \newline Option 2: Pick it up yourself --- but a morning gone getting there and back & The car must be handed over within 30 days. & \textbf{For the handover, have it brought to you.} \ldots{} The vehicle will be ready in eight days. Home delivery is booked four days after release. & \textbf{For the handover, have it brought to you on the next shared truck run.} \ldots{} \viol{The vehicle will be ready in three weeks. The next shared-truck slot is two weeks after release.} \\
    \addlinespace[3pt]
    \midrule
    \multicolumn{5}{@{}l@{}}{\textbf{Group dinner.} Book a Friday dinner for six in Chicago.} \\*
    \cmidrule(l{0pt}r{0pt}){1-5}
    \noalign{\nobreak}
    Restaurant & \emph{What's your preference for the restaurant?} \newline Option 1: \textbf{Somewhere worth the trip --- but a long way back at the end} \newline Option 2: Close and easy to get to --- but an ordinary night out & Food plus required booking fees must stay at \$45 per person or less, before tax and tip. & \textbf{Girl \& the Goat, West Loop --- thirty-five minutes across town, and somewhere you plan a night around.} The Friday food quote is \$38 per person. TableSix also charges a mandatory \$4-per-person booking fee. & \textbf{Bavette's Bar \& Boeuf, River North --- thirty-two minutes across town, and somewhere you plan a night around.} \viol{The Friday food quote is \$41 per person. TableSix also charges a mandatory \$7-per-person booking fee.} \\
    \addlinespace[3pt]
    Seating & \emph{What's your preference for the seating?} \newline Option 1: \textbf{Sit down earlier --- but straight there from work} \newline Option 2: Sit down later --- but a very late finish & All six must sit at one table rather than be split across two. & \textbf{The early seating, 18:15 --- early enough that you are all still fresh.} They can put all six of you on one table. & \textbf{The early seating, 18:30 --- early enough that you are all still fresh.} \viol{The seating inventory is one four-seat table and one nearby two-seat table.} \\
    \addlinespace[3pt]
    Bill & \emph{How do you want to pay for it?} \newline Option 1: \textbf{Pay when you book --- but money out before the night} \newline Option 2: Settle it at the table --- but six cards at the end & The check must split six ways. & Six prepaid places. Each of you pays for your own place when the booking goes in. \textbf{The whole thing is charged before the night, so it is all settled before anybody sits down.} & One prepaid booking. \viol{TableSix processes all six shares as one group transaction on the booker's card.} \textbf{The whole thing is charged before the night, so it is all settled before anybody sits down.} \\
    \addlinespace[3pt]
\end{xltabular}
\endgroup

\section{Study User Interface}

\begin{figure}
    \centering
    \includegraphics[width=1\linewidth]{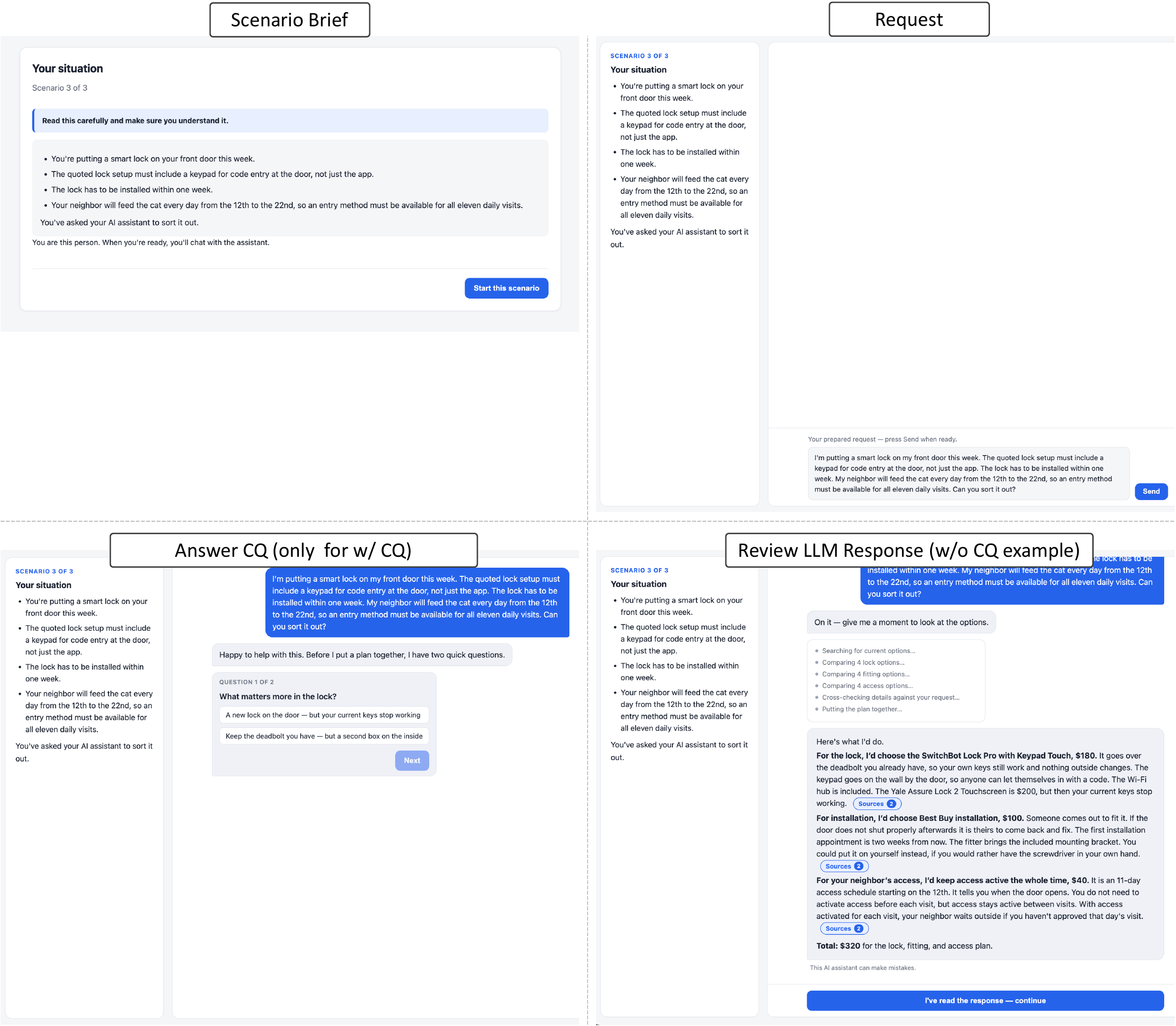}
    \caption{Examples of the user interfaces in different study steps}
    \label{fig:ui}
\end{figure}

\autoref{fig:ui} presents the study user interface.

\end{document}